\documentclass[aps,pra,twocolumn,showpacs,groupedaddress]{revtex4}
\usepackage{graphicx} 
\usepackage{amsmath}
\usepackage{longtable}

\begin{document}

\title{Positron-atom scattering using pseudo-state energy shifts}
\author{J.Mitroy}
  \email{jxm107@rsphysse.anu.edu.au}
\affiliation{ARC Center for Anti-Matter Studies, Faculty of Technology, Charles Darwin University, Darwin NT 0909, Australia}
\author{J.Y.Zhang}
\affiliation{ARC Center for Anti-Matter Studies, Faculty of Technology, Charles Darwin University, Darwin NT 0909, Australia}
\author{M.W.J.Bromley}
  \email{mbromley@physics.sdsu.edu}
\affiliation{Department of Physics and Computational Science Research Center, San Diego State University, San Diego CA 92182, USA}
\author{S.I.Young}
\affiliation{Department of Physics and Computational Science Research Center, San Diego State University, San Diego CA 92182, USA}

\date{\today}

\begin{abstract}

A method to generate low-energy phase shifts for elastic  
scattering using bound-state calculations is applied to the problem 
of $e^+$-Mg and $e^+$-Zn scattering after an initial validation
on the $e^+$-Cu system.  The energy shift between a small reference 
calculation and the largest possible configuration interaction 
calculation of the lowest energy pseudo-state is used to tune a 
semi-empirical optical potential.   
The potential was further fine-tuned by utilizing the energy
of the second lowest pseudo-state.  The $s$- and $p$-wave 
phase shifts for positron scattering from Mg and Zn are given from 
threshold to the first excitation threshold.  The $e^+$-Mg cross 
section has a prominent $p$-wave shape resonance at an energy of 
about 0.096 eV with a width of 0.106 eV.  The peak cross section 
for $e^+$-Mg scattering is about 4800 $a_0^2$ while $Z_{\rm eff}$ 
achieves a value of 1310 at an energy of 0.109 eV. 

\end{abstract}


\maketitle 

One of the most technically demanding problems in quantum physics
is the scattering problem, i.e. the prediction of the reaction 
probabilities when two objects collide \cite{burke95}.  
The underlying difficulty lies in the unbounded nature of 
the wave function.  This leads to a variety of computational and 
analytic complications that are absent in bound state 
calculations, e.g. the Schwartz singularities that occur in 
the Kohn variational method for scattering 
\cite{schwartz61a,nesbet80a}.    

One approach to solve scattering problems is to use methods 
that have been used for bound state calculations
\cite{burke95,schneider88a,bromley03a}.  There are many examples 
of such approaches, one of the most popular being the $R$-matrix 
methods that use the solutions of the Schr{\"o}dinger equation in a 
finite sized cavity to determine the behavior of the wave function 
in the interaction region \cite{fano73a,burke71a,barrett83a,burke95}.  
The total wave function is constructed by splicing the inner wave 
function onto the asymptotic wave function.           

This article had its origin in a particular scattering problem, 
namely the determination of the near threshold phase shifts for 
positron scattering from the di-valent group II and IIB atoms. The
dimension of the secular equations for bound state calculations 
on such systems are very large, for example a configuration 
interaction (CI) calculation of 
the $e^+$Ca $^2$P$^{\rm o}$ state resulted in equations of
dimension 874,448 \cite{bromley06f}.  These dimensions are much 
larger than those that would occur in a CI calculation of the 
$^2$P$^{\rm o}$ ground state Ca$^-$.  The exceptionally large
dimensionalities occur because the valence electrons tend to 
localize around the positron, thus giving a very slowly convergent 
partial wave expansion of the wave function  
\cite{mceachran77,higgins90,strasburger95,schrader98,mitroy99c,dzuba99}.  
The ability to routinely solve the secular equations associated with 
the CI basis using iterative sparse matrix techniques  
\cite{stathopolous94a} is one reason why CI calculations for positronic 
atoms (and of course for molecular systems) have been able to generate 
useful results.  

Trying to generate scattering solutions for such systems would
be problematic for a number of reasons.  For example, application 
of the CI-Kohn approach \cite{bromley03a} to determine the phase 
shifts for positron scattering from any group II or IIB atom
would result in linear equations with dimensions between 400,000 
to 1,000,000.  These are simply too large to be solved by direct 
methods.  Iterative methods for large linear systems do exist, but 
there are no robust methods that absolutely guarantee convergence 
\cite{saad00a}. The development of an efficient linear solver for 
the class of problems that arise from a basis set treatment of 
quantum scattering would likely involve a good deal of initial 
experimentation and effort.  Similarly, the widely used $R$-matrix 
method with fixed boundary conditions \cite{burke71a} requires the 
generation of all the eigenvectors and eigenvalues of the Hamiltonian, 
which is not feasible when the matrix dimensions exceed 100,000.  

Very recently, a method was developed to extract phase shifts from the 
positive energies of a pair of CI calculations \cite{mitroy07a}.  In that 
work, the energy shifts of a positive energy pseudo-state were used to 
tune a semi-empirical optical potential which was then used to predict the 
close to threshold phase shifts.  This concept is refined in the present 
work and it is shown that the reliability of the potential can be 
enhanced by tuning the potential to the energies of the two lowest 
states.  Next, the $s$- and $p$-wave phase shifts for 
positron scattering from Mg and Zn are computed from threshold 
to the opening of the lowest energy excitation channel.  The Mg 
and Zn atoms are interesting for positron scattering experiments 
since Mg has been recently shown to possess a prominent $p$-wave 
shape resonance \cite{mitroy07a}.  The Zn system is also interesting 
since the existence of a $e^+$Zn bound state of $^2$S$^{\rm e}$ 
symmetry \cite{mitroy99a} will manifest itself in a differential 
cross section that is largest at backward angles \cite{bromley02d}.     
   
\section{Model independent method for generating phase shifts} 

\subsection{The box variational method} 

The idea behind the current method lies closest to the box variational 
method \cite{risberg56a,percival57a,percival60a} which is exploited 
in quantum Monte Carlo (QMC) calculations of scattering 
\cite{alhassid84a,carlson84a,shumway01a,chiesa02a,nollet07a}.  In the 
box variational method, 
one extracts the phase shift by comparing the zero point energy of 
a finite size cavity to the energy of the system wave function in 
the same cavity. In its simplest incarnation for $s$-wave scattering, 
one diagonalizes the Hamiltonian, in natural units where $\hbar = m = e = 1$,
\begin{equation}
H = - \frac{1}{2} \nabla^2 + V(r)  , 
\label{hamV} 
\end{equation}
in a cavity of radius $R$.  The wave function obeys the boundary
conditions $\Psi(0) = \Psi(R) = 0$. The positive energy states
$\Phi_n(r)$, with energy $E_n$, can be regarded as the small $r$-part 
of the exact scattering wave function, $\Psi_n(r)$, with that same energy.  
The exact wave function can be written as $\Psi_n(r) = \sin(k_n r +\delta)$ 
for $r > R$ where $\delta$ is the phase shift and the wavenumber $k_n = \sqrt{2E_n}$.
At the boundary, one has $\sin(k_n R + \delta) = 0$, giving 
\begin{equation}
\delta = n \pi - k_n R ,  
\label{sphase} 
\end{equation}
(this expression assumes there are no bound states).          

For systems with non-zero angular momentum, the asymptotic $\sin$-wave
is replaced by the asymptotic form 
$\psi(r) \sim j_\ell(kr) + \tan(\delta_\ell) n_k(kr)$ 
where $j_\ell(kr)$ and $n_\ell(kr)$ are spherical Bessel functions
of the 1st and 2nd kind.  The condition $\psi(R) = 0$ gives the
following expression for the phase shift,
\begin{equation}
\tan( \delta_\ell(k_n)) = - \frac{j_\ell(k_nR)}{n_\ell(k_nR)} \ .  
\label{lphase} 
\end{equation}

\subsection{Phase shifts using pseudo-state energy shifts.} 

The box variational method has two advantages, (a) it is very 
simple to apply and (b) the $B$-spline basis sets currently in use 
in many atomic structure applications easily satisfy the
necessary boundary conditions.  However, there are other basis 
sets in use that do not satisfy the $\Psi(0) = \Psi(R) = 0$ 
boundary conditions.  

Consider the usage of a set of general $L^2$ functions, $u = \{ \phi_i \}$.  
These functions have a finite radial extent and thus the basis 
can be regarded as defining a soft-sided cavity.  A simple procedure
is used here to estimate the radius of the resulting soft-sided box.
Denoting $E[u,0]_n$ to be the $n$th energy eigenstate resulting from 
a diagonalization of the $V = 0$ potential in the basis $u$,
then the effective radius of the soft box is given by  
\begin{equation}
R = \frac{X_{\ell n}}{\sqrt{2E[u,0]_n}} = \frac{X_{\ell n}}{k[u,0]_n} \ ,  
\label{Rbasis} 
\end{equation}
where $X_{\ell n}$ is the $n$th zero of the spherical Bessel 
function, $j_\ell(x)$.

The potential $V = V(r)$ is then diagonalized in the same
basis to give $E[u,V]_n$, and hence $k[u,V]_n = \sqrt{2E[u,V]_n}$.    
The phase shift is then extracted using
\begin{equation}
\tan( \delta_\ell) = - \frac{j_\ell(k[u,V]_n R)}{n_\ell(k[u,V]_n R)} \ .
\label{Rphase} 
\end{equation}

\begin{figure}[tbh]
\centering{
\includegraphics[width=8.4cm,angle=0]{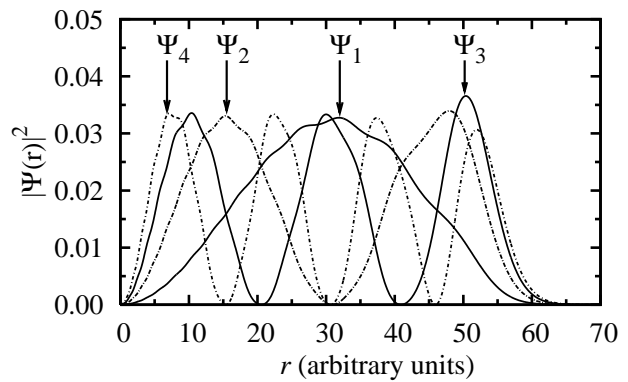}
}
\caption[]{ \label{freewave}
The radial probability densities, $|\Psi(r)|^2$,
for the four lowest $s$-wave pseudo-states resulting from a
diagonalization of the $V = 0$ potential in a basis of 30 LTOs.
The pseudo-states have all been normalized to unity.
}
\end{figure}

Figure \ref{freewave} shows the radial probability density,
$|\Psi(r)|^2 = |\sum_\alpha c_\alpha r\chi_{\alpha}|^2$,
of the four lowest energy $\ell=0$ wave functions computed by the diagonalization of
a $V = 0$ potential in a basis of 30 Laguerre Type Orbitals (LTOs) with the scale 
parameter $\lambda = 1$. The general definition for the LTOs were
\begin{equation}
 \chi_\alpha(r)=N_\alpha r^{\ell} \exp (-\lambda _\alpha r)
L_{n_\alpha -\ell - 1}^{(2\ell +2)}(2\lambda _\alpha r) \ ,
\label{LTO}
\end{equation}
where the normalization constant is
\begin{equation}
N_\alpha =\sqrt{\frac{(2\lambda_\alpha) (n_\alpha-\ell-1)!}
{(\ell+n_\alpha+1)!}} \ .
\label{LTOnorm}
\end{equation}
The function $L_{n_\alpha-\ell-1}^{(2\ell +2)}(2 \lambda _\alpha r)$ is
an associated Laguerre polynomial that can be defined in terms of a
confluent hyper-geometric function \cite{bromley02a}.
The probability densities go to zero for $r > 60$.  The wiggles
in the probability densities are not a numerical artifact, rather 
they are a manifestation of the slow convergence of the $L^2$ basis 
to the exact continuum wave function \cite{stelbovics90a}.    

Table \ref{softbox} gives the energies, and the effective radius,  
of the soft box for this basis as given by eq.~(\ref{Rbasis}).  
The four states have an effective radius of about 61.5 which is
consistent with Figure \ref{freewave}.  It is reasonable to
conclude that eq.~(\ref{Rbasis}) gives an estimate of the
range of a pseudo-state that is sufficiently accurate to be
useful.    

This LTO basis was also used to diagonalize the Woods-Saxon 
potential, 
\begin{equation}
V(r) = - \frac{V_0}{1 + \exp\left(\frac{(r - W_0)}{a} \right)} \ ,
\label{woodssaxon}
\end{equation}
with the choice $V_0 = 0.97$, $W_0 = 1$ and $a = 0.05$.
The energies and phase shifts derived from eq.~(\ref{Rphase}) 
are listed in Table \ref{softbox}.  The phase shifts 
obtained by numerically integrating the Schr{\"o}dinger equation
for the Woods-Saxon potential are also listed in Table 
\ref{softbox} and are exact to all quoted
digits.  The two sets of phase shifts agree with each other
to an accuracy of about 2$\%$.   

\begin{table}[th]
\caption[]{  \label{softbox}
Parameters derived from the diagonalization of the free-wave
and Woods-Saxon potential in a basis of 30 LTOs with
$\lambda = 1$.  The pseudo-state energies for $V = 0$ is 
denoted $E_{0}$ while the Woods-Saxon energies are denoted 
$E_{\rm ws}$.  The radius of the soft-box is denoted $R_0$,
while the phase shift from eq.~(\ref{Rphase}) is $\delta$.
The phase shift obtained by integrating the Woods-Saxon
potential numerically is $\delta_{\rm exact}$.      
}
\begin{ruledtabular}
\begin{tabular}{lccccc} 
   $n$ &  $E_{0}$ & $R_0$ & $E_{\rm ws}$ & $\delta$ & $\delta_{\rm exact}$ \\ \hline
\multicolumn{6}{c}{$\ell = 0$ }   \\    
    1  & 0.001286  & 61.95 &  0.001170 & 0.1451 & 0.1470  \\  
    2  & 0.005170  & 61.79 &  0.004705 & 0.2897 & 0.2852  \\
    3  & 0.011734  & 61.52 &  0.010736 & 0.4098 & 0.4092  \\  
    4  & 0.021116  & 61.15 &  0.019362 & 0.5333 & 0.5148  \\ \hline  
\multicolumn{6}{c}{$\ell = 1$ }   \\    
    1  & 0.002474  & 63.87 &  0.002450 & 0.0212 & 0.0211  \\  
    2  & 0.007361  & 63.67 &  0.007147 & 0.1114 & 0.1103  \\
    3  & 0.014810  & 63.36 &  0.013964 & 0.3132 & 0.3073  \\  
    4  & 0.024969  & 62.94 &  0.022844 & 0.6087 & 0.5903 \\   
\end{tabular}
\end{ruledtabular}
\end{table}

\begin{figure}[tbh]
\centering{
\includegraphics[width=8.4cm,angle=0]{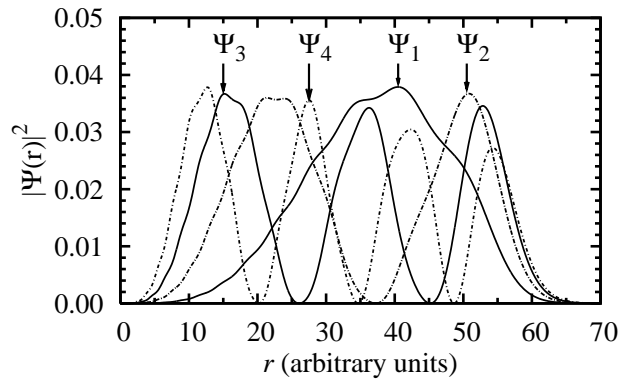}
}
\caption[]{ \label{freewave1}
The radial probability densities, $|\Psi(r)|^2$,
for the four lowest $p$-wave pseudo-states resulting from
a diagonalization of the $V = 0$ potential in a basis of 30 LTOs.
The pseudo-states have all been normalized to unity.
}
\end{figure}

This procedure has also been validated for $p$-wave scattering.
Figure \ref{freewave1} shows the result of diagonalizing the
$V = 0$ potential for $p$-wave scattering in a basis of 30 LTOs
with $\ell = 1$ and $\lambda = 1.0$.  Once again the range of
the pseudo-state solutions are roughly the same.  Table 
\ref{softbox} lists the effective box radius for each pseudo-state 
as derived from eq.~(\ref{Rbasis}). 

The Woods-Saxon potential with the choice $V_0 = 0.173$, $W_0 = 4.0$
and $a = 1.0$ was then diagonalized in this basis.  The phase shifts
obtained from  eq.~(\ref{Rphase}) are tabulated in Table 
\ref{softbox} along with phase shifts generated by a numerical
solution of the Schr{\"o}dinger equation.  The two sets of phase 
shifts agree to within 3$\%$.    

\section{Model dependent method for generating phase shifts} 

It has long been known that the positive energy pseudo-states resulting 
from the diagonalization of the Hamiltonian in an $L^2$ basis 
often give a reasonable approximation to the exact scattering
wave function over a finite range  
\cite{harris67a,hazi70a,drachman75a,drachman76b,ivanov01b,ivanov02c}.  
Methods which exploit this result are sometimes called stabilization 
methods.  While convergence of the pseudo-state to 
the continuum wave function is relatively slow, matrix elements 
formed by the pseudo-state often have reasonable convergence 
properties \cite{stelbovics90a}.  In effect, while the point-wise 
properties of the wave function can be inaccurate, the convergence 
in the mean of the wave function over a suitable range can be 
quite good.  This raises the possibility that the expectation values
of positive energy pseudo-states can be used to define a semi-empirical
optical potential to describe low-energy scattering.     

Our method proceeds as follows. The initial calculation uses a 
reference basis of square integrable single particle orbitals, 
$\{ \phi_i({\mathbf r}) \}$, designed to give a good representation
of the wave function in a bounded interaction region.  The 
Hamiltonian, $H_0$ for a free particle with $V = 0$ is diagonalized, 
yielding the wave function 
\begin{equation}
\Phi_0  = \sum_i c_i \phi_{i}({\mathbf r}) \ ,  
\label{refwffree}
\end{equation}
and the energy expectation,  
\begin{equation}
E_{\rm free} = \langle \Phi_0 |  H_{0} | \Phi_0 \rangle .   
\label{refEfree}
\end{equation}

The wave function of the target atom is constructed from a linear 
combination of configurations $\{ \omega_{i}({\mathbf X}) \}$, 
of the same symmetry as the ground state (${\mathbf X}$ is the 
collective set of target coordinates).  So one can write 
\begin{equation}
\Omega_{\rm gs}({\mathbf X}) = \sum_i c_i \omega_{i}({\mathbf X})  \ ,  
\label{refwf2}
\end{equation}
while
\begin{equation}
E_{\rm gs} = \langle \Omega_{\rm gs} |  H_{\rm target} | \Omega_{\rm gs} \rangle .   
\label{refEgs}
\end{equation}
The reference energy, $E_0$, is determined by diagonalizing the Hamiltonian 
in the product basis,  ${\Omega_{\rm gs}({\mathbf X}) \phi_{i}({\mathbf r})}$,  
to give 
\begin{equation}
E_{0} = \langle \Psi_0 |  H_{\rm target} | \Psi_0 \rangle \ ,    
\label{refE0}
\end{equation}
where 
\begin{equation}
\Psi_0  = \sum_i c_i \phi_{i}({\mathbf r}) \Omega_{\rm gs}({\mathbf X}) \ . 
\label{refwf0}
\end{equation}

The basis sets $\{ \phi_i({\mathbf r}) \}$ and 
$\{ \omega_{i}({\mathbf X}) \}$ are then augmented by a  
large number of additional functions 
$\{ \chi_i({\mathbf r}) \}$ and $\{ \psi_{i}({\mathbf X}) \}$ 
to represent the
correlations between the projectile and the target constituents.  
None of these additional functions have the same symmetries 
as those used in $\{ \phi_i({\mathbf r}) \}$ and 
$\{ \omega_{i}({\mathbf X}) \}$.  This augmented trial function
can be written as 
\begin{equation}
\Psi_1  = \sum_{i,j} c_{i,j} \omega_{\rm i}({\mathbf X})  \phi_{j}({\mathbf r})  
        + \sum_{i,j} d_{i,j} \psi_{\rm i}({\mathbf X})  \chi_{j}({\mathbf r}) \ . 
\label{exactwf}
\end{equation}
The trial wave function, $\Psi_1$, is used to diagonalize $H_{\rm exact}$
giving an energy     
\begin{equation}
E_1 = \langle \Psi_{1} |  H_{\rm exact} | \Psi_{1} \rangle .      
\label{exactE}
\end{equation}

Next, the basis $\{ \phi_i({\mathbf r}) \}$ is diagonalized in  
a parameterized potential designed to describe the most  
important features of the interaction between the projectile 
and the target.   This potential can be written formally as 
\begin{equation}
V_{\rm opt}(r)  = V_{\rm dir}(r) + V_{\rm pol}(r)  \ .  
\label{vopt}
\end{equation}
The potential $V_{\rm dir}$ is the direct interaction between the
target and projectile.  This can be approximated by the direct 
interaction between the projectile and the target Hartree-Fock 
(HF) ground state wave function, $\Omega_{\rm HF}$, which can
be slightly
different from $\Omega_{\rm gs}$. The polarization potential
$V_{\rm pol}(r)$ is semi-empirical in nature with the asymptotic
form  
\begin{equation}
V_{\rm pol}(r)  \sim - \frac{\alpha_d}{2r^4}  \ ,  
\label{vpol0}
\end{equation}
where $\alpha_d$ is the static dipole polarizability of the target.  
In previous work, \cite{bromley98,mitroy02a,bromley02d,mitroy07a} 
a simple one-parameter form 
\begin{equation}
V_{\rm p1}(r)  = - \frac{\alpha_d}{2r^4}  \left( 1 - \exp(-r^6/\rho^6 ) \right) ,  
\label{vpol1}
\end{equation}
has usually been adopted for $V_{\rm pol}(r)$.  It is thought that this 
functional form has the incorrect shape at intermediate values of $r$, 
e.g. $r \approx 5$ $a_0$ (The reasons why we originally become suspicious
about the reliability of eq.~(\ref{vpol1}) are not discussed here.  
But the results obtained later will clearly show the limitations of 
this type of cutoff polarization potential).  The present work will also 
use a more complicated expression 
for $V_{\rm pol}(r)$, with an additional adjustable parameter, $A_Q$
to give an improved description of the potential between target and atom.  
This form was    
\begin{eqnarray}
V_{\rm p2}(r)  &=& - \frac{\alpha_d}{2r^4}  \left( 1 - \exp(-r^6/\rho^6 ) \right) \nonumber  \\  
               &-& \frac{A_Q}{2r^6}  \left( 1 - \exp(-r^8/\rho^8 ) \right) .  
\label{vpol2}
\end{eqnarray}
While the second term has the functional form of a quadrupole
polarization potential, it should be regarded as primarily empirical
in nature.  This functional form was chosen as a screened quadrupole     
type potential because it was computationally convenient.  

The energy expectation value of the ground state, or lowest energy
pseudo-state     
\begin{equation}
E_{\rm opt} = \langle \Phi_{\rm opt} |  H_{\rm opt} | \Phi_{\rm opt} \rangle ,      
\label{optE}
\end{equation}
is adjusted by tuning the parameters in $V_{\rm p1}$ until 
$E_{\rm opt} = E_{1} - E_{\rm gs} $.  

Determination of the $V_{\rm p2}$ required additional information
since there are two parameters, $\rho$ and $A_Q$ that need to be
fixed.  In this case, the optical potential is tuned to 
two energy levels rather than one.  This does increase the
overall time of the calculation since it is necessary to extract 
the lowest two eigenvalues from the CI 
calculation.

Once the optical potential has been fixed, it is a simple matter 
to generate the exact continuum solution of the Schr{\"o}dinger 
equation for the Hamiltonian given by eq.~(\ref{vopt}).   

\subsection{Positron annihilation}   

Besides obtaining the phase shifts in the low-energy region, it is also possible to determine 
the annihilation parameter, $Z_{\rm eff}$ \cite{fraser68a,mceachran77,ryzhikh00a}.
The fundamental idea is to compare exact and model potential calculations
of $Z_{\rm eff}$, and so fix the enhancement factor, $G$ \cite{mitroy02a,mitroy02g,mitroy05f}.  
Enhancement factors were first introduced in the calculation of
the annihilation rate of positrons in condensed matter systems
\cite{boronski86,puska94,barbiellini01a}.  
They incorporate the tendency for attractive electron-positron 
correlations to increase the electron density at the position of 
the positron. 

It has been shown that model potential calculations of $s$-wave positron 
scattering from hydrogen and helium that were tuned to give the correct 
phase-shift at a reference energy also reproduced the low-energy
behavior of $Z_{\rm eff}(k)$ up to a multiplying constant 
(i.e. $G$) \cite{mitroy02a}.  
The annihilation parameter for the model potential wave function 
follows the model of Mitroy and Ivanov \cite{mitroy02a}, and is written as 
\begin{equation}
Z_{\rm eff} =  \int d^3r 
       \Bigl( G_v\rho_v({\bf r}) + G_c\rho_c({\bf r}) \Bigr) 
             \left| \Phi_{\rm opt}({\bf r})\right|^2 \  , 
            \label{Zeff}
\end{equation}
where $\rho_c({\bf r})$ and $\rho_v({\bf r})$ 
are the electron densities associated with the core and valence 
electrons of the target atom, and $\Phi_{\rm opt}({\bf r})$ is the 
positron scattering function obtained in the tuned model 
potential. The notation  $Z^{(\ell)}_{\rm eff}$ is used to
denote the annihilation parameter for the $\ell$th partial 
wave.  

For the core orbitals, $G_{c}$ is set to 2.5 due to reasons 
outlined in Ref.~\cite{mitroy02a}.  The valence enhancement factor 
$G_{v}$ is computed by the simple ratio
\begin{equation}
G_{v}  = \frac {\Gamma^{CI}_{v}}{\Gamma^{\rm model}_{v}}  \ ,  
\label{Gnorm} 
\end{equation}
where $\Gamma^{CI}_v$ is the annihilation rate of the positron
with the valence orbitals as given by the CI calculation and 
$\Gamma^{\rm model}_v$ is the valence annihilation rate predicted 
by the model potential calculation with $G = 1$.  

\section{The fixed core potentials} 

All calculations on the $e^+$Cu, $e^+$Mg, and $e^+$Zn 
systems used a fixed core Hamiltonian.  The details of the core
potentials have been discussed previously
\cite{bromley02a,bromley02b,bromley02d,bromley02e,mitroy06a}, 
but a short description is worthwhile.  
The model Hamiltonian is initially based on a 
HF wave function for the neutral atom ground state.
One and two-body semi-empirical polarization potentials are added 
to the potential field of the HF core and the parameters of 
the core-polarization potentials defined by reference to 
the spectra of Cu, Mg$^+$ and Zn$^+$ 
\cite{bromley02a,bromley02b,bromley02d,bromley02e}.

The effective Hamiltonian for the systems with 2 valence 
electrons (${\bf r}_1$ and ${\bf r}_2$) and a positron 
(${\bf r}_0$) was   
\begin{eqnarray}
H  &=&  - \frac{1}{2}\nabla_{0}^2 - \sum_{i=1}^{2} \frac {1}{2} \nabla_{i}^2 
- V_{\rm dir}({\bf r}_0) + V_{\rm cp1}({\bf r}_0) \nonumber \\  
&+& \sum_{i=1}^{2} (V_{\rm dir}({\bf r}_i) + V_{\rm exc}({\bf r}_i) + V_{\rm cp1}({\bf r}_i)) - \sum_{i=1}^{2} \frac{1}{r_{i0}} \nonumber \\  
   &+&  \frac{1}{r_{12}}  
   - V_{\rm cp2}({\bf r}_1,{\bf r}_2)
   + \sum_{i=1}^{2} V_{\rm cp2}({\bf r}_i,{\bf r}_0) \ .
\end{eqnarray}
The direct potential ($V_{\rm dir}$) represents the interaction 
with the HF electron core.  The direct part of the core potential 
is attractive for electrons and repulsive for the positron.  
The exchange potential  ($V_{\rm exc}$) between the valence 
electrons and the HF core was computed without approximation.

The one-body core polarization potentials ($V_{\rm cp1}$) are semi-empirical
in nature.  They have the functional form
\begin{equation}
V_{\rm cp1}(r)  =  -\sum_{\ell m} \frac{\alpha_d g_{\ell}^2(r)}{2 r^4} 
                    |\ell m \rangle \langle \ell m| .
\label{corepolar1}
\end{equation}
The factor $\alpha_d$ is the static dipole polarizability of
the core and $g_{\ell}^2(r)$ is a cut-off function designed to make 
the polarization potential finite at the origin.  The same 
cut-off function has been adopted for both the positron and 
electrons.  In this work, $g_{\ell}^2(r)$ was defined to be
\begin{equation}
g_{\ell}^2(r) = 1-\exp\bigl(-r^6/\rho_{\ell}^6 \bigr) \ ,
                                    \label{cut1} 
\end{equation}
where $\rho_{\ell}$ is an adjustable parameter.  
The two-body polarization potential ($V_{\rm cp2}$) is defined as
\begin{equation}
V_{\rm cp2}({\bf r}_i,{\bf r}_j) = \frac{\alpha_d} {r_i^3 r_j^3}
({\bf r}_i\cdot{\bf r}_j)g_{\rm cp2}(r_i)g_{\rm cp2}(r_j)\ . 
                                    \label{polar2}    
\end{equation}
where $g_{\rm cp2}(r)$ is chosen to have a cut-off parameter, $\rho_{\rm cp2}$,
obtained by averaging the $\rho_{\ell}$.  The core dipole polarizabilities were 
set to 0.4814 $a_0^3$ for Mg \cite{bromley02a,bromley02b}, 
5.36 $a_0^3$ for Cu \cite{bromley02e}, and 2.294 $a_0^3$ for Zn 
\cite{bromley02d}.  The cutoff parameters for Mg were 
$\rho_0 = 1.1795$ $a_0$, $\rho_1 = 1.302$ $a_0$, $\rho_2 = 1.442$ $a_0$,  
$\rho_{3} = 1.52$ $a_0$, $\rho_{\ell \ge 4} = 1.361$ $a_0$,
and $\rho_{\rm cp2} = 1.361$ $a_0$.
The cutoff parameters for Cu were 
$\rho_0 = 1.9883$ $a_0$, $\rho_1 = 2.03$ $a_0$, $\rho_2 = 1.83$ $a_0$,  
$\rho_{3} = 1.80$ $a_0$, $\rho_{\ell \ge 4} = 1.91$ $a_0$,
and $\rho_{\rm cp2} = 1.91$ $a_0$.
The cutoff parameters for Zn were 
$\rho_0 = 1.63$ $a_0$, $\rho_1 = 1.80$ $a_0$, $\rho_2 = 2.30$ $a_0$,  
$\rho_{3} = 1.60$ $a_0$, $\rho_{\ell \ge 4} = 1.83$ $a_0$,
and $\rho_{\rm cp2} = 1.83$ $a_0$.
This model has been used to describe many of the features of neutral  
Be, Mg, Ca and Sr to quite high accuracy 
\cite{bromley02a,bromley02b,mitroy03f}.

\section{Verification for $e^+$-C\lowercase{u} scattering} 

Previously, a validation of the method was performed for 
$s$-wave $e^+$-H scattering \cite{mitroy07a}.  In the 
present work the method is further verified by computing the low-energy
phase shifts and annihilation parameters for $s$-wave 
and $p$-wave $e^+$-Cu scattering.  The model copper 
atom used here has a dipole polarizability of $41.65$ $a_0^3$ \cite{bromley02a}
and, therefore, provides a more stringent test of the procedure
used to tune the shape of the polarization potential than 
the previous test upon the $e^+$-H system (where $\alpha_d = 4.5$ $a_0^3$).

The explicit CI calculation on the $e^+$-Cu ground state and the
CI-Kohn calculations of $e^+$-Cu scattering closely follow those 
previously reported \cite{bromley02e,bromley03a,mitroy06a}. 
Briefly, the wave function expansion consists of a large number
of single particle orbitals and includes terms with $\ell > 10$.  
The single particle orbitals are usually represented as  
Laguerre type orbitals (LTOs).     

The $e^+$Cu ground state calculation included orbitals up to $L \le 16$
with a minimum of 18 electron LTOs and 18 positron LTOs per $\ell$.
The CI reference wave function, $\Psi_0$, consisted of the copper 
atom ground state multiplied by a positron basis of 30 $\ell = 0$ 
LTOs.  The orbital basis was slightly reduced for the 
calculation of the lowest energy $^2$P$^o$ pseudo-state.  In 
this case, the calculation included terms up to $L = 14$ with 
a minimum of 18 electron LTOs and 18 positron LTOs per $\ell$.
The CI reference wave function, $\Psi_0$, in this case consisted 
of the copper atom ground state multiplied by a positron basis of 
33 $\ell = 1$ orbitals.

\begin{table}[th]
\caption[]{  \label{atomicdata}
Expectations values obtained from CI calculations of the 
bound $^2$S$^e$ states and the $^2$P$^o$ pseudo-states for
the $e^+$Cu, $e^+$Mg and $e^+$Zn systems.  The binding 
energy $\varepsilon$ (in Hartree) is negative for bound
states and positive for pseudo-states.  The mean positron radius
$\langle r_p \rangle$ is in units of $a_0$. The core ($\langle \Gamma \rangle_c$)
and valence $(\langle \Gamma_v \rangle)$ annihilation rates are
given in units of $10^9$ s$^{-1}$.  All of the values given in this
table are the results of extrapolating $L \to \infty$.
}
\begin{ruledtabular}
\begin{tabular}{lccccc}
System & Symmetry   &  $\varepsilon$ & $\langle r_p \rangle$ & $\langle \Gamma \rangle_c$ & $\langle \Gamma_v \rangle$  \\
\hline
$e^+$Cu &  $^2$S$^e$  & $-$0.005124  &  9.037 &  0.0322    & 0.5035  \\
$e^+$Cu &  $^2$P$^o$  & 0.001860   &  35.23 &  0.000413  & 0.0186  \\
$e^+$Mg &  $^2$S$^e$  & $-$0.01704   &  6.930 &  0.0109    & 1.004  \\
$e^+$Mg &  $^2$P$^o$  & 0.003989   &  13.87 &  0.00110   & 0.3729  \\
$e^+$Zn &  $^2$S$^e$  & $-$0.003794  &  9.726 &  0.0244    & 0.4269  \\
$e^+$Zn &  $^2$P$^o$  & 0.006885   &  20.24 &  0.000609  & 0.0190  \\
\end{tabular}
\end{ruledtabular}
\end{table}

One difficulty present in all CI calculations of positron-atom interactions 
is the slow convergence of the energy with $L$ \cite{mitroy99c,mitroy02b,mitroy06a}.  
The convergence pattern of the atomic CI expansion
\cite{schwartz62a,carroll79a,hill85a,salomonson89b,gribakin02a,mitroy06a,bromley07a}, 
suggests the use of an asymptotic analysis that utilizes the result that successive 
increments, $\Delta E_{L} = \langle E \rangle_L - \langle E \rangle_{L-1}$, 
can be written as an inverse power series, \textit{viz}.
\begin{equation}
\Delta E_L \approx \frac {A_E}{(L+{\scriptstyle \frac{1}{2}})^4} 
    + \frac {B_E}{(L+{\scriptstyle \frac{1}{2}})^5} 
    + \frac {C_E}{(L+{\scriptstyle \frac{1}{2}})^6} + \dots \ \   .
\label{extrap1}
\end{equation}
The $L \to \infty$ limits have been determined by fitting sets of 
$\langle E \rangle_L$ values to asymptotic series with either 1, 2 
or 3 terms.  The factors, $A_E$, $B_E$ and $C_E$ for the 
3-term expansion are determined at a particular $L$ from 4 successive 
energies ($\langle E \rangle_{L-3}$, $\langle E\rangle_{L-2}$, 
$\langle E \rangle_{L-1}$ and $\langle E \rangle_{L}$).  The series 
is summed to $\infty$ once the linear factors have been determined.  

Some expectation values of the $e^+$Cu $^2$S$^e$ ground state and 
the lowest energy $^2$P$^o$ pseudo-state in the $L \to \infty$
limits are given in Table \ref{atomicdata}.  It should be noted
that the leading term of the inverse power series for  
the annihilation rate, $\Gamma$, is 
$A_{\Gamma}/(L+1/2)^2$ \cite{gribakin02a,mitroy06a}.
There is some uncertainty in the extrapolation procedure and we 
estimate uncertainties of about 1$\%$ for the energy and 5$\%$ 
for the annihilation rate.  However, this does not impact the
present verification exercise.  The extrapolation 
procedures were applied consistently to both the CI calculations used 
to define the model potentials (and enhancement factors), as well as the 
independent CI-Kohn scattering calculations \cite{bromley03a} used to validate 
the model potential calculations.  Note that the errors in the extrapolated
results introduced by the use of a finite basis set have a tendency
to fortuitously cancel out \cite{mitroy06a}.

The trial function, $\Psi_0$, was then used to diagonalize the model 
potential, eq.~(\ref{vpol2}) with two different polarization 
potentials.  In the first instance, eq.~(\ref{vpol1}) was used 
and the parameter $\rho$ varied until the energy matched that
of the CI calculation.  This potential will be referred to 
as the $V_{\rm p1}$ potential.  In the second instance, the 
parameters, $A_Q$ and $\rho$ of eq.~(\ref{vpol2}) 
were both varied until both
the energy of the ground state and lowest energy pseudo-state 
were the same as the CI calculations.  
This potential will be termed the $V_{\rm p2}$ potential.  
The enhancement factor, $G_v$, was determined after the model
potentials were finalized.  In the case of the $V_{\rm p2}$
potential the ratio in eq.~(\ref{Gnorm}) was evaluated for 
the ground state.  The details of the model potential  
parameters are summarized in Table \ref{cumgznpot}.  

\begin{figure}[tbh]
\centering{
\includegraphics[width=8.4cm,angle=0]{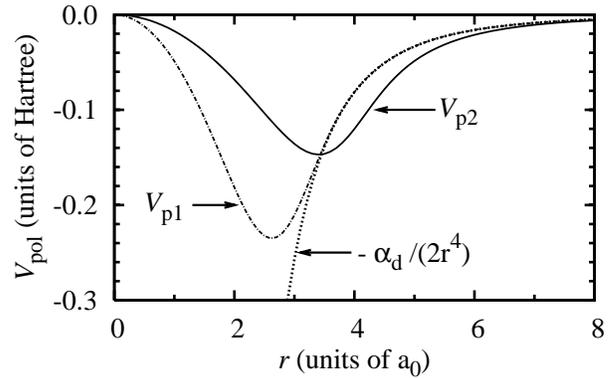}
}
\caption[]{ \label{Vpol}
Comparison of the two different parameterizations of the $e^+$-Cu 
$^2$S$^e$ polarization potential of eq.~(\ref{vpol1}) and
eq.~(\ref{vpol2}) against that of the asymptotic form of eq.~(\ref{vpol0}).
}
\end{figure}

\begin{table*}[th]
\caption[]{  \label{cumgznpot}
Definitions of the $V_{\rm p1}$ and $V_{p2}$ model potentials used to 
describe $s$-wave and $p$-wave scattering of $e^+$-Cu, $e^+$-Mg and 
$e^+$-Zn.  The $s$-wave potentials were tuned to the properties 
of the $^2$S$^e$ ground state and the lowest energy pseudo-state,
while the $p$-wave potentials were tuned to the two lowest energy $p$-wave pseudo-states.
The binding energy $\varepsilon$ (in Hartree) is negative for bound
states and positive for pseudo-states.  The mean positron radius
and scattering length, $A_{\rm scat}$, are in units of $a_0$. 
The core and valence annihilation rates are
given in units of $10^9$ s$^{-1}$.
}
\begin{ruledtabular}
\begin{tabular}{lccccccccccc} 
Atom & Potential &  $L$    &  $\alpha_d$ & $A_Q$ & $\rho$ & $G_v$ & $\varepsilon$   & 
$\langle r_p \rangle $    &  $\langle \Gamma_c \rangle$ & $\langle \Gamma_v \rangle$  &  $A_{\rm scat}$  \\ \hline
Cu &  $V_{\rm p1}$ &  0 & 41.65 &  0.0   &  2.7434 &  18.94  & $-$0.005124  &  8.46  &  0.0730  & 0.5036 &   12.4  \\
Cu & $V_{\rm p2}$ &  0 & 41.65 &  480.0  & 3.6248 &  26.35 & $-$0.005124  &  8.822 & 0.04088   & 0.5034 & 12.8 \\ 
Cu & $V_{\rm p1}$ &   1 & 41.65 &  0.0   &  2.1231 &  20.01 &  0.0057801 &  35.18  &  0.00275 & 0.0186  &     \\
Cu & $V_{\rm p2}$ &  1 & 41.65 &  360.0  & 3.0829 &  36.70 &  0.0057801 &  35.22  & 0.000868 & 0.0186 &     \\
Mg & $V_{\rm p1}$ &   0 & 71.35 &  0.0   &  2.9927 &  13.12  & $-$0.017072   &  6.21  &  0.0243   & 1.004  &  6.09    \\
Mg & $V_{\rm p2}$ &  0 & 71.35 & 2280.0  & 4.4794 &  24.74 & $-$0.017072   &  6.982 &  0.00738   & 1.004  &  7.23  \\
Mg & $V_{\rm p1}$ &   1 & 71.35 &  0.0   &  2.5626 &  12.35  & 0.003989   &  12.90 &  0.00654   & 0.3729 & \\
Mg & $V_{\rm p2}$ &  1 & 71.35 & 1250.0  & 3.8406 &  28.15  & 0.003989   &  13.80 &  0.00115  & 0.3729 &     \\
Zn & $V_{\rm p1}$ &   0 & 41.25 &  0.0   &  2.6579 &  9.91  & $-$0.003794  &  9.34  &  0.0412   & 0.4269  &  14.3   \\
Zn & $V_{\rm p2}$ &  0 & 41.25 &  430.0  & 3.5344 &  14.35  & $-$0.003794  &  9.71  &  0.0219   & 0.4269  & 14.7   \\
Zn &  $V_{\rm p1}$ &  1 & 41.25 &  0.0   &  2.1604 &  10.45 & 0.006885   &  20.21 &  0.00177  & 0.0190 &  \\
Zn & $V_{\rm p2}$ &  1 & 41.25 &  252.0  & 3.0117 &  17.45 & 0.006885   &  20.26 &  0.000770 & 0.0190 &    \\
\end{tabular}
\end{ruledtabular}
\end{table*}

Figure \ref{Vpol} shows a comparison of the $V_{\rm p1}$ 
and $V_{\rm p2}$ polarization potentials for the $^2$S$^e$ 
symmetry.  The $V_{\rm p1}$ potential is always smaller 
in magnitude than the $\alpha_d/(2r^4)$ asymptotic form.
The $V_{\rm p2}$ potential bulges below the $\alpha_d/(2r^4)$ 
asymptotic form and is stronger than a pure dipole potential
in the outer valence region of the atom.  This is entirely
reasonable.  The slow convergence of the single-center 
expansion occurs as a result of the localization of the 
valence electrons in the vicinity of positron
\cite{mceachran77,higgins90,strasburger95,schrader98,mitroy99c,dzuba99}.  
This, in turn, enhances the strength of the polarization 
potential in the outer valence region.  

The superiority of the $V_{\rm p2}$ potential in describing the
the $^2$S$^e$ bound state is apparent from Tables 
\ref{atomicdata} and \ref{cumgznpot}.  The $V_{\rm p2}$ calculation 
overestimates the core annihilation rate by 26$\%$ while the
$V_{\rm p1}$ potential overestimates this parameter by
more than 120$\%$.  Additionally, the $V_{\rm p2}$ potential
gives a better estimate of the mean positron radius,
$\langle r_p \rangle$.  The value of 8.822 $a_0$ is about 
2$\%$ smaller than the CI value of 9.037 $a_0$ while 
$V_{\rm p1}$ gave $\langle r_p \rangle = 8.46$ $a_0$ ($\approx 6\%$ smaller).

Accurate phase shifts for the full $e^+$-Cu scattering 
Hamiltonian were obtained from CI-Kohn variational calculations 
\cite{bromley03a} of the $e^+$-Cu system using {\em exactly} the
same short-range orbital basis sets as used in the CI calculation.  
The only difference between the CI-Kohn and regular CI basis sets 
is the addition of two continuum basis functions \cite{bromley03a}.  
The phase shifts for the $V_{\rm p1}$ and $V_{\rm p2}$ potentials 
were obtained by integrating the Schr{\"o}dinger equation.

\begin{figure}[tbh]
\centering{
\includegraphics[width=8.4cm,angle=0]{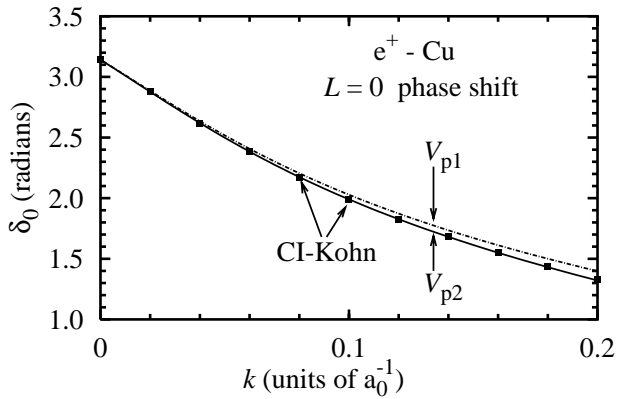}
}
\caption[]{ \label{Cuphases}
The phase shift for $e^+$-Cu scattering in the $s$-wave as a 
function of $k$ (in units of $a_0^{-1}$).  The lines 
show the results of the present calculation using the 
tuned $V_{p1}$ and $V_{p2}$ potentials while the  
squares show the phase shifts of the explicit CI-Kohn 
calculation.
}
\end{figure}

The $V_{\rm p2}$ scattering length estimate of 12.8 $a_0$ 
is within 2$\%$ of the CI-Kohn estimate of the scattering length, 
namely $13.05$ $a_0$.  The $V_{\rm p1}$ scattering length of $12.4$ 
$a_0$ is too small by 5$\%$.  Figure \ref{Cuphases} shows the 
comparison between the model potential $s$-wave phase shift and 
the CI-Kohn phase shift for $k \in [0,0.2]$ $a_0^{-1}$.  The 
$V_{\rm p1}$ model slightly overestimates 
the CI-Kohn phase shifts (modulo $\pi$) over the entire range.  
The $V_{\rm p2}$ fit to the CI-Kohn phase shifts is clearly 
superior.  

\begin{figure}[tbh]
\centering{
\includegraphics[width=8.4cm,angle=0]{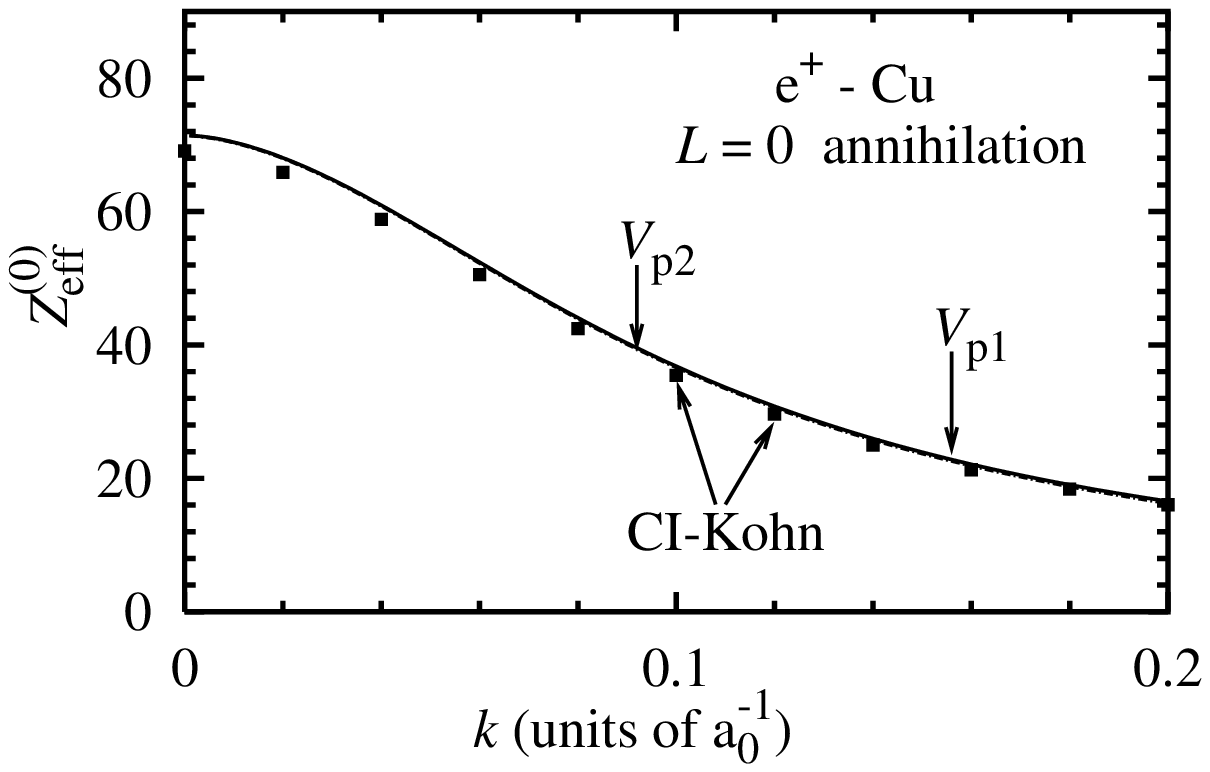}
}
\caption[]{ \label{cuzeff0}
The annihilation parameter, $Z^{(0)}_{\rm eff}$ for $e^+$-Cu 
$s$-wave scattering as a function of $k$ (in units of $a_0^{-1}$).  
The two curves were calculated with the $V_{\rm p1}$ and 
$V_{\rm p2}$ potentials.  The discrete points are taken from
the explicit CI-Kohn calculations.   
}
\end{figure}

\begin{figure}[tbh]
\centering{
\includegraphics[width=8.4cm,angle=0]{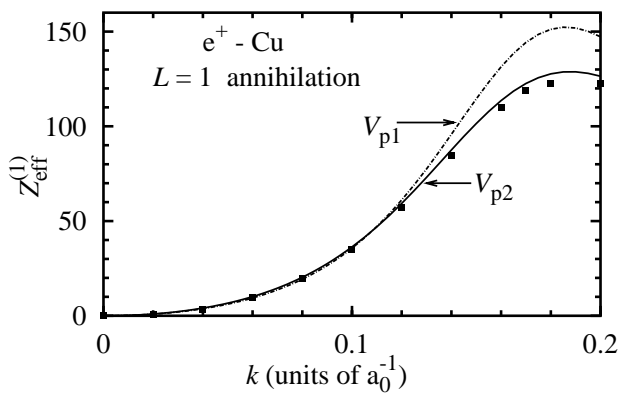}
}
\caption[]{ \label{cuzeff1}
The annihilation parameter for $p$-wave scattering, $Z^{(1)}_{\rm eff}$ for 
$e^+$-Cu scattering as a function of $k$ (in units of $a_0^{-1}$).  
The different curves were calculated with the $V_{\rm p1}$ and 
$V_{\rm p2}$ potentials.  The discrete points are taken from
the explicit CI-Kohn calculations.   
}
\end{figure}

Besides obtaining phase shifts, this procedure was used to determine 
the valence annihilation parameter which is shown in Figures 
\ref{cuzeff0} and \ref{cuzeff1}.  The $V_{\rm p2}$ enhancement 
factor of $G_v = 26.35$, gives an $s$-wave annihilation parameter,
$Z^{(0)}_{\rm eff}$, that is within 5$\%$ of the 
explicit CI-Kohn calculation over the entire energy 
range.  Somewhat surprisingly, the $Z^{(0)}_{\rm eff}$ from the
$V_{\rm p1}$ model is almost the same as that from the
$V_{\rm p2}$ model. 

\begin{figure}[tbh]
\centering{
\includegraphics[width=8.4cm,angle=0]{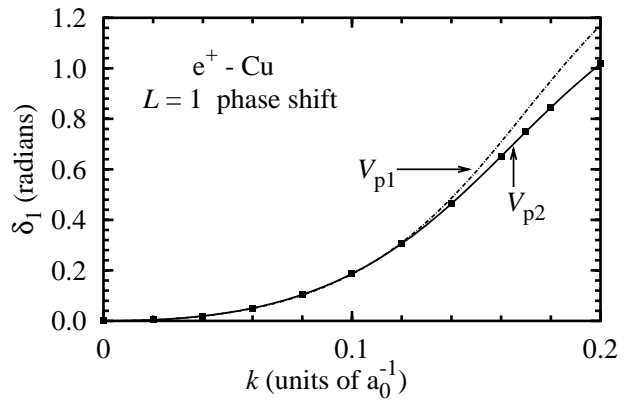}
}
\caption[]{ \label{Cuphasep}
The phase shift for $e^+$-Cu scattering in the $p$-wave as a 
function of $k$ (in units of $a_0^{-1}$).  The lines 
show the results of the present calculation using the 
tuned $V_{p1}$ and $V_{p2}$ potentials while the squares 
show the phase shifts of the explicit CI-Kohn 
calculation.
}
\end{figure}

The phase shift for $p$-wave scattering is shown in Figure \ref{Cuphasep}.
The $V_{\rm p1}$ potential overestimates the CI-Kohn phase shift as the energy
increases and there is a 15$\%$ discrepancy at $k = 0.20$ $a_0^{-1}$.  The
$V_{\rm p1}$ potential also tends to underestimate the phase shift 
for $k < 0.10$ $a_0^{-1}$, although this is difficult to see from the
Figure.  The $V_{\rm p2}$ potential reproduces the CI-Kohn phase shifts
very well and the agreement is perfect within the resolution of the graph.

This pattern is repeated in Figure \ref{cuzeff1} where $Z^{(1)}_{\rm eff}$ 
is plotted as a function of $k$.  The $V_{\rm p1}$ potential tends to
overestimate the CI-Kohn values at the higher momenta with the discrepancy 
at $k = 0.20$ $a_0^{-1}$ being 15$\%$.  However, the $V_{\rm p2}$ potential 
does an excellent job of reproducing the CI-Kohn $Z^{(1)}_{\rm eff}$ over
the entire momentum range.  The $V_{\rm p2}$ $Z^{(1)}_{\rm eff}$ is too
large at the higher momentum, but the difference is only $2\%$ at
$k = 0.20$ $a_0^{-1}$.

The final confirmation of the improved quality of the $V_{\rm p2}$ 
potential comes from the comparisons of the pseudo-state expectation values 
in Tables \ref{atomicdata} and \ref{cumgznpot}.  The $\langle r_p \rangle$ 
given by $V_{\rm p2}$ is closer to the CI value than the $V_{\rm p1}$ 
value.  Furthermore, the $V_{\rm p2}$ potential is better than the 
$V_{\rm p1}$ potential at reproducing the CI core annihilation rate of  
$0.0322 \times 10^{9}$ $s^{-1}$ (this value assumes $G_c = 1$).     

\section{Positron scattering from magnesium} 

\subsection{The CI calculations}

Although many of the specifics of the calculations upon $e^+$Mg 
have been reported previously \cite{bromley06c,mitroy07a}, 
further details concerning the wave function construction are given 
here.   The trial wave function adopted for the CI calculations
consists of a linear combination of states which are anti-symmetric 
in the interchange of the two electrons,   
\begin{equation}
|\Psi;LS \rangle_a  =  \sum_{i} c_{i} |\Phi_{i};LS\rangle_a \ . 
\label{kohnwf}
\end{equation}
Each anti-symmetrized state is constructed as a linear
combination of coupled but not anti-symmetrized states.
Two electrons (particles 1 and 2) are coupled first to 
each other, then the positron (particle 0) is coupled to 
form a state with net angular and spin angular momentum, 
$L$ and $S$.  The anti-symmetric states are written as
\begin{eqnarray}
|\Phi_i; [a b] L_I S_I p LS\rangle_a \! & \! = &\! \! \frac{1}{\sqrt{2(1+\delta_{ab})}}
\Bigl( \  |[a_1 b_2]L_I S_I p_0 \rangle \nonumber \\  
   & \! + & \! \! (-1)^{\ell_a+\ell_b+L_I+S_I} |[a_2 b_1 ]L_I S_I p_0 \rangle \!  \Bigr) \ , \nonumber \\ 
\end{eqnarray}
where the subscript by each orbital denotes the electron
occupying that particular orbital.  

The $e^+$Mg CI basis was constructed by letting the two electrons and 
the positron form all the possible configuration with a total angular 
momentum of $L$, with the two electrons in a spin-singlet 
state, subject to three selection rules,
\begin{eqnarray}
\max(\ell_0,\ell_1,\ell_2) & \le & L \ , \\
\min(\ell_1,\ell_2)& \le & L_{\rm int} \ ,  \\  
(-1)^{(\ell_0+\ell_1+\ell_2)}& \equiv & +1 \; \mathrm{or} -1 \ . 
\end{eqnarray}
In these rules $\ell_0$, $\ell_1$ and $\ell_2$ are respectively 
the orbital angular momenta of the positron and the two electrons.
The even [odd] parity states require $(-1)^{(\ell_0+\ell_1+\ell_2)} \equiv +1$ [$-1$].

The Hamiltonian for the $e^+$Mg $^2$S$^{\rm e}$ state was
diagonalized in a CI basis including orbitals up to $\ell = 12$.
There were a 
minimum of $15$ radial basis functions for each $\ell$.  There 
were 19 $\ell = 0$ positron orbitals.  The largest $^2$S$^{\rm e}$ calculation 
was performed with $L = 12$ and $L_{\rm int} = 4$.  The 
$L_{\rm int}$ parameter does not have to be large since it is 
mainly concerned with describing the more quickly converging 
electron-electron correlations \cite{bromley02b}.    
The CI basis for the $e^+$Mg $^2$P$^{\rm o}$ symmetry included 
orbitals up to $\ell = 14$.  There were a 
minimum of $14$ radial basis functions for each $\ell$.  There 
were 20 $\ell = 1$ positron orbitals.  The largest $^2$P$^{\rm o}$ calculation 
was performed with $L = 14$ and $L_{\rm int} = 3$.  

A summary of $e^+$Mg expectation values taken to the 
$L \to \infty$ limit are given in Table \ref{atomicdata}.  
The binding energy $\varepsilon$ for each symmetry is calculated 
with respect to the energy of the Mg ground state using the
basis for that symmetry.  The overall binding energy of
the $^2$S$^e$ ground state was $-0.017072$ Hartree, with
the first pseudo-state at 0.002503 Hartree.  The energies of the
two lowest pseudo-states of $^2$P$^o$ symmetry were 0.003989 and
0.012012 Hartree respectively.
 
\subsection{Model potential calculations} 

\begin{figure}[tbh]
\centering{
\includegraphics[width=8.4cm,angle=0]{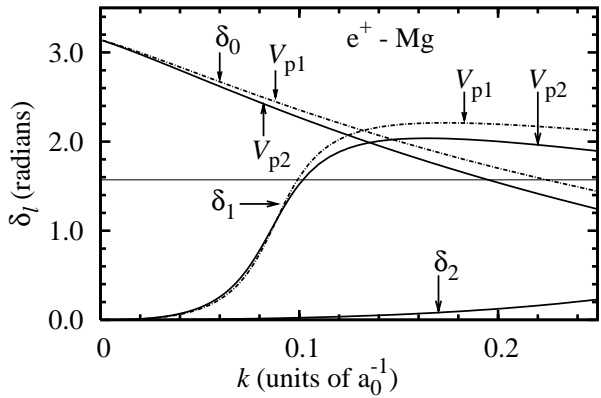}
}
\caption[]{ \label{mgphase}
The $s$, $p$ and $d$-wave phase shifts for $e^+$-Mg scattering as a 
function of $k$ (in units of $a_0^{-1}$).  The solid lines were 
computed with $V_{\rm p2}$ while the dashed lines were computed
with $V_{\rm p1}$.  The horizontal line shows $\delta = \pi/2$.  
}
\end{figure}

The $\rho$ parameter of the $V_{\rm p1}$ potential was tuned to reproduce
the energies of the lowest $^2$S$^e$ and $^2$P$^o$ states.  The values 
of $\rho$ and the expectation values of the lowest state of each symmetry 
are given in Table \ref{cumgznpot}.   The values of $A_Q$ and $\rho$ for 
the $V_{\rm p2}$ potential  were tuned to the lowest two energies.  
Examination of the expectation values of Tables \ref{atomicdata} and 
\ref{cumgznpot} reveals that the $V_{\rm p2}$ potential again does a 
better job at reproducing the CI expectation values.  The $V_{\rm p1}$ 
potential underestimates the mean positron radius by 10$\%$ and further 
overestimates the core annihilation rate by a factor of 2.  The 
$V_{\rm p2}$ potential gives a value of $\langle r_p \rangle$ that is 
too large by 1$\%$.  The $V_{\rm p2}$ underestimation of $\Gamma_c$ is 
about 30$\%$.  While the $V_{\rm p2}$ 
model potential may not be perfect, it does a better job of 
describing the radial distribution of the positron density 
than the $V_{\rm p1}$ potential.

\begin{figure}[tbh]
\centering{
\includegraphics[width=8.7cm,angle=0]{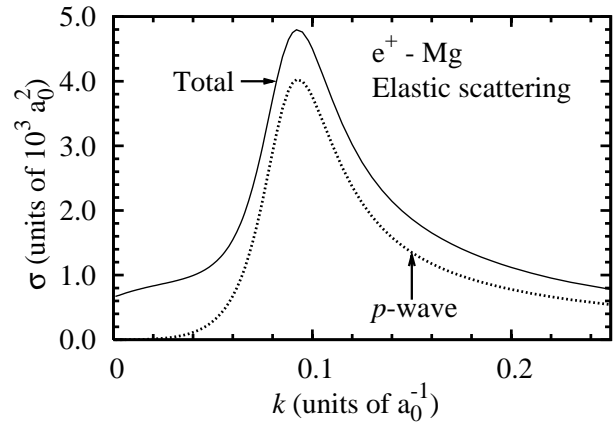}
}
\caption[]{ \label{mgcross}
The elastic scattering cross section for $e^+$-Mg scattering 
as calculated with the $V_{\rm p2}$ potential 
in the energy region below the Ps-formation threshold at
$k \approx 0.249$ $a_0^{-1}$.  The 
solid line shows the total cross section while the dashed 
curve shows the $\ell = 1$ partial cross section.   
}
\end{figure}

The situation for the $^2$P$^o$ pseudo-state is similar to
that for the $^2$S$^e$ state.  The $V_{\rm p1}$ potential 
underestimates the mean positron radius by 10$\%$ and 
overestimates the core annihilation rate by a factor of 6.  
The $V_{\rm p2}$ potential on the other hand gives an
$\langle r_p \rangle$ within 1$\%$ of the CI value and  
overestimates the core annihilation rate by only 5$\%$.

The $s$- and $p$-wave phase shifts are plotted in Figure 
\ref{mgphase}.  The 10$\%$ difference between the two
model potential scattering lengths manifests itself in
the slightly different $s$-wave phase shifts.  The 
difference between the $V_{\rm p1}$ and $V_{\rm p2}$ 
potentials is larger for the $p$-wave phase shift, although
both predict a resonance at $k \approx 0.09$ $a_0^{-1}$.  The 
$d$-wave phase shift plotted in Figure \ref{mgphase} 
was computed with the $V_{\rm p2}$ $p$-wave potential.  The 
$\ell > 2$ phase shifts used in the computation of the total 
cross section also used the $V_{\rm p2}$ $p$-wave potential.    
 
Figure \ref{mgcross} shows the elastic cross section for $e^+$-Mg 
scattering below the Ps formation threshold (at $k \approx 0.25 \ a_0^{-1}$)   
as computed with the $V_{\rm p2}$ potentials.  The $p$-wave resonance 
leads to the total elastic cross section achieving a peak value of 
4800 $a_0^2$.

\begin{figure}[tbh]
\centering{
\includegraphics[width=8.4cm,angle=0]{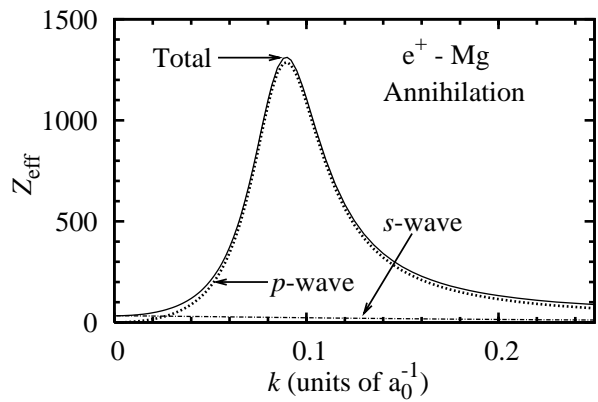}
}
\caption[]{ \label{mgzeff}
The annihilation parameter $Z_{\rm eff}$ for $e^+$-Mg scattering as 
a function of $k$ (in units of $a_0^{-1}$).  The different curves
show, $Z^{(0)}_{\rm eff}$, $Z^{(1)}_{\rm eff}$ and 
$Z^{({\rm total})}_{\rm eff}$ as calculated with the $V_{\rm p2}$ 
potential.   
}
\end{figure}

The existence of the resonance in the phase shift also leads to
a resonance in the annihilation parameter, $Z_{\rm eff}$  
\cite{mitroy07a}.  The curve in Figure \ref{mgzeff} was computed
with an enhancement factor of $G_c = 2.5$ for core annihilation, 
an $s$-wave valence enhancement factor of $G_v = 24.7$ and a 
$p$-wave enhancement factor of $G_v = 28.2$.  The total
$Z_{\rm eff}$ shown in Figure \ref{mgzeff} is almost completely
dominated by the contribution from the $p$-wave and the   
value of $Z_{\rm eff}$ at the resonance peak was 1310.

The resonance parameters were determined by performing a 
fit to the function
\begin{equation}
\delta = \delta_0 + a(E-\varepsilon_R) + 
         \tan^{-1}\left( \frac{\Gamma}{2(\varepsilon_R - E)} \right)
         + bE^2 \ .   
\end{equation}
This gave a value of $\varepsilon_R = 0.00351$ Hartree for the
resonance position and a width of $\Gamma = 0.00390$ Hartree.   

\subsection{Reliability of the resonance prediction}

The possible sources of error in the $e^+$-Mg calculations 
are (a) the reliability of the underlying model potential
for the CI calculation,
(b) the extent to which the CI calculations have 
converged and (c) the ability of the scattering model 
potential to reliably reproduce the scattering parameters.

The cutoff parameter for the positron part of the core polarization 
potential (refer to eq.~(\ref{corepolar1})) is chosen to be the same as 
the cutoff for the electron.  This is likely to underestimate the strength 
of the positron interaction since there is a good deal of evidence for 
closed shell systems that suggests the positronic part of the polarization 
potential is stronger than the electronic part \cite{mitroy02b}.   However, 
the impact of this is likely to be small since the core polarizability 
of 0.4814 $a_0^3$ is more than 100 times smaller than the neutral atom 
polarizability of 71.35 $a_0^3$.  Any correction would tend to shift
the resonance to a lower energy and increase the height of the 
maximum in the cross section.  

The CI calculations of the $^2$S$^e$ ground state are believed 
to be converged to about 2$\%$ in the energy.  An independent 
calculation of the $e^+$Mg ground state has been done with the 
fixed core stochastic variational  method (FCSVM) \cite{mitroy08f}.  
The FCSVM Hamiltonian is very similar to the fixed core Hamiltonian 
used for the present calculation and the current best FCSVM estimate of 
the binding energy is 0.017117 Hartree.  
However, it has also proved possible to make an estimate 
of the variational limit of the FCSVM calculation.  This 
estimate is between 0.01735 and 0.01740 Hartree 
\cite{mitroy08f} which is about 2$\%$ more tightly bound 
than the CI calculation.  The calculation of the $^2$P$^o$ 
state is expected to have an accuracy similar to that of 
the ground state.

The existence and position of the resonance is independent of 
the exact form of $V_{\rm pol}$.  Besides the calculations reported 
here, alternate calculations with some other parameterizations
were reported earlier \cite{mitroy07a}.  All of these calculations
gave a resonance almost at the same position and magnitude.  
The reason for this lies in the accident that the energy of 
the $^2$P$^o$ pseudo-state, at $k = 0.0893$ $a_0^{-1}$,
lies close to the center of the resonance. 
At this energy, the determination of the phase shift will
be largely model independent since the stabilization concept
ensures that the $L^2$ wave function is a reasonable 
approximation to the actual continuum wave function.    
The phase shifts of the two potentials at the energy of
the pseudo-state were $\delta_1 = 1.157$ and $\delta_1 = 1.153$ radians,
for $V_{\rm p1}$ and $V_{\rm p2}$ respectively.  Additional plots of the phase shifts obtained
with other functional forms for $V_{\rm pol}$ have tended
to have a common intersection point near $k \approx 0.089$ $a_0^{-1}$.

Finally, the simple potential independent approach of 
eq.~(\ref{Rphase}) has been applied to determine the phase
shift at the pseudo-state energy.  The energy of the 
positron $p$-wave LTO basis in the $V = 0$ potential 
was 0.007572 Hartree.  The radius of the box giving this
energy is $R_{\rm box} = 36.5$ $a_0$.  Evaluating 
eq.~(\ref{Rphase}) at $k = 0.0893$ $a_0^{-1}$ gives  
$\delta = 1.16$ radian.
 
\section{Positron scattering from zinc} 

Positron binding to zinc has been known with some degree of certainty 
since 1999 \cite{mitroy99a,bromley02d} following some earlier, less 
conclusive, work \cite{szmytkowski93b,dzuba95b,mceachran98}.  
The neutral zinc atom has an ionization potential of 0.34523 Hartree
\cite{nistasd312} and a polarizability of $38.8\pm0.8$ $a_0^3$
\cite{goebel96b}.  The present model potential for the Zn$^{2+}$ 
core predicts an ionization potential of $0.33519$ Hartree and a
polarizability of 41.25 $a_0^3$ \cite{bromley02d}.  

The present CI calculations upon the $e^+$Zn ground state used the 
same core potential as the earlier CI calculations
\cite{bromley02d}, but 
the size of the basis has been enlarged.  The 
maximum number of partial waves has been increased to $L = 12$, 
the number of LTOs per $\ell$ has been increased to 16, and
finally $L_{\rm int}$ was increased from 3 to 4.  The overall
dimension of the CI calculation has increased by an order of
magnitude.  The summary of $e^+$Zn expectation values for the 
series of calculations with increasing $L$ are given in Table 
\ref{tab:zn}.  The energy of the Zn ground state with respect
to the Zn$^{2+}$ core for the electron basis was $-0.99549251$ 
Hartree.  
 
\begin{table*}[th]
\caption[]{  \label{tab:zn}
Results of the CI calculations for $e^+$Zn atoms for a given 
$L$.  The $E$ column gives the energy with 
respect to the doubly ionized frozen core and $\varepsilon$ is the
binding energy with respect to the lowest energy dissociation 
channel at $E = -0.99549251$ Hartree.  The radial expectation
values (in $a_0$) of the electron and positron are listed in 
the $\langle r_e \rangle$ and $\langle r_p \rangle$ columns.
The $\langle \Gamma_v \rangle$ and $\langle \Gamma_c \rangle$ 
columns give the valence and core annihilation rates (in 
$10^9$ sec$^{-1}$).  The results in the row labeled 10* 
are taken from an earlier calculation \cite{bromley02d}.  The 
results under the heading $L \to \infty$ incorporate an 
$L \rightarrow \infty$ correction.  
}
\begin{ruledtabular}
\begin{tabular}{lccccccccc} 
$L$&  $N_e$ & $N_p$ & $N_{CI}$ & $\langle E \rangle_L$ & $\langle \varepsilon \rangle_L$ & $\langle r_e \rangle_L$ & $\langle r_p \rangle_L$   & $\langle \Gamma_c \rangle_L$ & $\langle \Gamma_v \rangle_L$  \\ \hline   
 0 &  19  &  16   &  3040  &  $-$0.97217702  &  $-$0.02331549  &  2.76525 &  29.69360 &  0.0002583 & 0.0002144 \\ 
 1 &  37  &  32   &  11248 &  $-$0.99240348  &  $-$0.00308903  &  2.75421 &  26.43204 &  0.0009532 & 0.0023477 \\ 
 2 &  55  &  48   &  30112 &  $-$0.99441912  &  $-$0.00107339  &  2.75879 &  21.51407 &  0.0037086 & 0.0144975 \\ 
 3 &  71  &  64   &  58336 &  $-$0.99562488  &  0.00013237   &  2.77148 &  16.80037 &  0.0086564 & 0.0456435 \\
 4 &  87  &  80   &  101264 & $-$0.99653983  &  0.00104732   &  2.78651 &  13.89104 &  0.0133632 & 0.0872535 \\
 5 &  103 &  96   &  153744 & $-$0.99719181  &  0.00169930   &  2.79963 &  12.32951 &  0.0167251 & 0.1279350 \\
 6 &  119 &  112  &  210576 & $-$0.99768537  &  0.00219287   &  2.80986 &  11.46661 &  0.0189393 & 0.1632760 \\
 7 &  135 &  128  &  271248 & $-$0.99805166  &  0.00255915   &  2.81770 &  10.94916 &  0.0204117 & 0.1929861 \\
 8 &  151 &  144  &  334096 & $-$0.99832262  &  0.00283011   &  2.82371 &  10.61913 &  0.0214113 & 0.2176885 \\
 9 &  167 &  160  &  398864 & $-$0.99852427  &  0.00303176   &  2.82833 &  10.39734 &  0.0221079 & 0.2382207  \\
 10 & 183 &  176  &  463632 & $-$0.99867583  &  0.00318336   &  2.83190 &  10.24428 &  0.0226023 & 0.2553248  \\
 11 & 199 &  192  &  528400 & $-$0.99879107  &  0.00329856   &  2.83471 &  10.13275 &  0.0229647 & 0.2696997 \\
 12 & 215 &  208  &  593168 & $-$0.99887979  &  0.00338728   &  2.83692 &  10.05093 &  0.0232336 & 0.2818534 \\
 10* \cite{bromley02d}& 104 &  97   &  63712  & $-$0.9983995   &  0.0030385    &  2.82927 &  10.32455 &  0.022292 & 0.24023  \\
\multicolumn{10}{c}{$L \to \infty$ extrapolations} \\ 
\multicolumn{4}{l}{Present} & $-$0.9992869   & 0.0037944  & 2.8475 &   9.72595 & 0.02434       & 0.42692     \\    
\multicolumn{4}{l}{Previous \cite{bromley02d}} & $-$0.999092   & 0.003731  & 2.8451 &   9.9139 & 0.02393       & 0.3927     \\    
\end{tabular}
\end{ruledtabular}
\end{table*}

An examination of Table \ref{tab:zn} reveals that the present extrapolated
binding energy of 0.0037944 Hartree is within 2$\%$ of the previously obtained 
binding energy.  To a certain extent, this high level of agreement is fortuitous.
The method used to extrapolate the energy increment to the $L \to \infty$
limit in Ref. \cite{bromley02d} had an inherent tendency to overestimate the
binding energy.  However, this compensated for the tendency of a finite
dimension LTO basis to increasingly underestimate the energy increment
as $L \to \infty$ \cite{mitroy06a}.  The lowest positive energy
$^2$S$^e$ pseudo-state had an energy of 0.0024706 Hartree above the Zn ground state.

\begin{figure}[tbh]
\centering{
\includegraphics[width=8.4cm,angle=0]{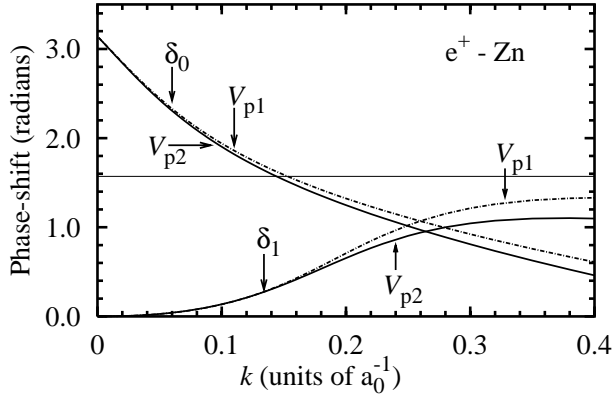}
}
\caption[]{ \label{znphase}
The $s$- and $p$-wave phase shifts for $e^+$-Zn scattering as a 
function of $k$ (in units of $a_0^{-1}$).  
The solid lines were 
computed with $V_{\rm p2}$ while the dashed lines were computed
with $V_{\rm p1}$.  
The horizontal dashed line shows $\delta = \pi/2$.  
}
\end{figure}

The Hamiltonian for the $e^+$Zn $^2$P$^{\rm o}$ state was
diagonalized in a CI basis including orbitals up to $\ell = 10$.  
The two electrons were in a spin-singlet state, with a
minimum of $16$ radial basis functions for each $\ell$.  There
were 20 $\ell = 1$ positron orbitals.  The largest calculation
was performed with $L = 10$ and $L_{\rm int} = 3$.  The energy of 
the lowest energy $^2$P$^o$ pseudo-state was 0.006885 Hartree above 
threshold.  Other expectation values for this state are listed
in Table \ref{atomicdata}.   The second lowest $^2$P$^o$ pseudo-state 
was located at 0.019055 Hartree.  

\begin{figure}[tbh]
\centering{
\includegraphics[width=8.7cm,angle=0]{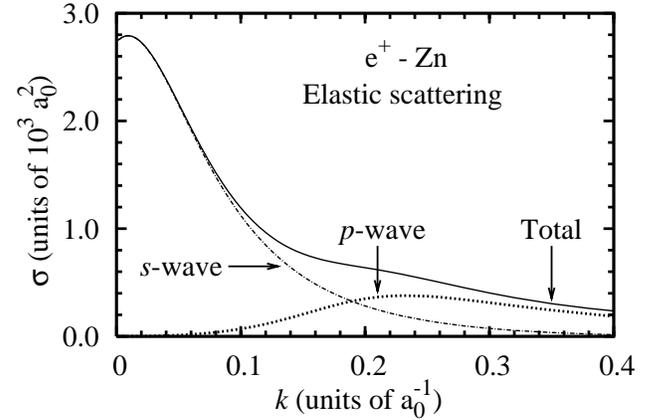}
}
\caption[]{ \label{zncross}
The elastic scattering cross section for $e^+$-Zn scattering 
as calculated with the $V_{\rm p2}$ potential 
in the energy region below the Ps-formation threshold at
$k \approx 0.436$ $a_0^{-1}$.  The 
solid line shows the total cross section while the two dashed 
curves show the $\ell = 0$ and $\ell = 1$ partial cross sections.   
}
\end{figure}

The parameters for the $V_{\rm p1}$ and $V_{\rm p2}$ potentials, 
tuned to the CI data in Table \ref{atomicdata}, are listed in
Table \ref{cumgznpot}.  A casual glance at the entries in these 
two Tables reveals that the $V_{\rm p2}$ potential again does  
better at reproducing the properties of the $^2$S$^e$ physical 
state and the $^2$P$^o$ pseudo-state.  

\begin{figure}[tbh]
\centering{
\includegraphics[width=8.4cm,angle=0]{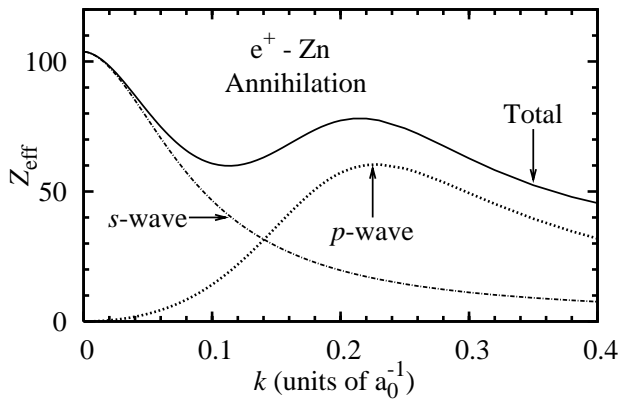}
}
\caption[]{ \label{znzeff}
The annihilation parameter $Z_{\rm eff}$ for $e^+$-Zn scattering as 
a function of $k$ (in units of $a_0^{-1}$).  The different curves
show, $Z^{(0)}_{\rm eff}$, $Z^{(1)}_{\rm eff}$ and 
$Z^{({\rm total})}_{\rm eff}$ as calculated with the $V_{\rm p2}$ 
potential 
}
\end{figure}

The $s$- and $p$-wave phase shifts are plotted in Figure \ref{znphase}. 
The small $e^+$Zn binding energy leads to a large value for the 
scattering length, namely $14.7 \pm 0.1$ $a_0$.  The energy region 
below $k \le 0.144 a_0^{-1}$ would be interesting for an experimental
investigation as the $s$-wave phase shifts are in a different
quadrant from all the other phase shifts.  Consequently, 
the differential cross section will be larger at backward 
angles than at forward angles.  This peaking of the near
threshold differential cross section at backward angles is
a signature of the existence of the $e^+$Zn bound state
\cite{bromley02d}. 

The low-energy elastic cross section below the Ps formation threshold 
as shown in Figure \ref{zncross} was computed with the $V_{\rm p2}$ 
potentials.  The phase shifts for $\ell \ge 1$ are taken from the 
$\ell = 1$ model potential.  The large value of the cross section at 
$E = 0$ is characteristic of a potential supporting a weak bound state.
The quickly rising $p$-wave phase shifts leads to a shoulder in
the cross section near $k \approx 0.2$ $a_0^{-1}$.  The $p$-wave causes 
a more pronounced structure in $Z_{\rm eff}$ which is easily noticeable 
in figure \ref{znzeff} as the bump at $k \approx 0.2$ $a_0^{-1}$.  

The experimental observation of the $e^+$Zn $p$-wave resonance precursor 
in the cross section would be complicated by the large 
$s$-wave cross section which tends to obscure the feature 
in the total elastic cross section.  The resonance structure 
would be most visible in a measurement of $Z_{\rm eff}$ or in
a differential cross section.         

\section{Comparisons with previous work} 

The present calculations are not the only calculations of 
the $e^+$Mg and $e^+$Zn scattering systems.  However, the 
other calculations were of a much more speculative nature 
\cite{kurtz81,szmytkowski93a,gribakin96,campeanu98a,mceachran98,bromley98,mitroy02a,peng07a}.  
For example, the many body perturbation theory-based calculation 
of Gribakin and King predicted that the $^2$P$^o$ symmetry 
of $e^+$Mg had a bound state \cite{gribakin96}.  None of the 
other calculations on the $e^+$-Mg system gave a cross section 
with the prominent $^2$P$^o$ shape resonance. 

While some previous model potential calculations were based on 
reasonable estimates of the $e^+$Mg binding energy 
\cite{bromley98,mitroy02a}, the
uncertainties in defining the functional form of the 
polarization potential detracted from the reliability 
of the $p$-wave phase shift.  The present calculations are
more definitive, and the main source of uncertainty is
in the definition of the underlying core polarization 
potential in the CI calculations.   

\section{Summary} 

A new technique has been used to determine the phase shifts 
for low-energy positron-atom scattering from magnesium and 
zinc.  The phase shifts are determined by tuning an optical 
potential to the energy of a bound state or a positive energy 
state.   The tuning of an optical potential to features such 
as bound state energies and resonance positions is well known.  
Tuning an optical potential to a pseudo-state energy shift is 
novel \cite{mitroy07a}.  One improvement over our previous 
implementation is the use of a second energy to fine-tune the 
shape of the optical potential.  Another possible improvement
requiring further research would be to use other expectation
values (e.g. the mean positron radius) to further refine the
shape of the optical potential.  The use of experimental 
information to tune optical potentials is known (e.g. the role 
of the deuteron radius in tuning the $n$-$p$ potential).

There are two different concepts that can be regarded as providing
motivation for the present approach.  The first is the stabilization
concept, namely, a positive energy pseudo-state will provide a
reasonable approximation to the scattering state with that energy
over a finite radial range \cite{harris67a,hazi70a}. The alternate
motivation comes from the box variational method, namely that the 
energy shift of the wave function in a hard-sided box is used to 
estimate the phase shift \cite{risberg56a,percival57a,percival60a}.
Diagonalization of the Hamiltonian in a finite dimension LTO basis    
can be regarded as equivalent to diagonalizing the Hamiltonian in 
a soft-sided box.

\begin{figure}[tbh]
\centering{
\includegraphics[width=8.7cm,angle=0]{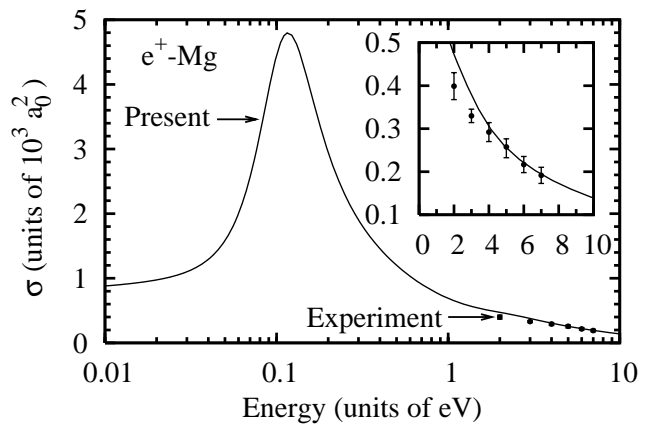}
}
\caption[]{ \label{Mgcrossv3}
The cross sections (in $10^3$ $a_0^2$) for $e^+$-Mg scattering versus energy (in eV).
The present elastic scattering cross sections as calculated with
the $V_{\rm p2}$ potential are shown as the solid line.
The total cross section measurements from the Detroit group 
are shown as the discrete points with their error bars \cite{stein98a}.
The inset shows the same data, plotted on a linear energy scale.
}
\end{figure}

The most significant result of the present investigation is the 
prediction of a close to threshold $^2$P$^o$ shape resonance
for elastic scattering from magnesium.  There is no experimental
evidence for the existence of shape resonances in positron-atom 
or positron-molecule scattering \cite{mitroy02b}. The 
present prediction has the virtue of being readily amenable 
to experimental verification.   Indeed, the Detroit group has
measured the total cross section for positron-magnesium
scattering down to an energy of 2.0 eV \cite{stein98a}.
As can be seen from Figure \ref{Mgcrossv3}, the lowest 
energy for which their measurements were done is just too 
high to detect the resonance.  An earlier experiment
measured down to an energy of 1.0 eV \cite{stein96}, but these
results are not shown in the figure as they are similar
to those in Ref. \cite{stein98a} whilst having larger reported errors.
Their most recent $e^+$-Mg measurements went down to $0.12$ eV,
however, they only reported the positronium formation
cross section \cite{surdutovich03a}.
  
With some reflection on the 
differences between positron-atom and electron-atom interaction 
potentials it is not surprising that the magnesium atom supports 
a shape resonance.  It has been noticed that the positrons 
are more strongly attracted to closed (sub)shell atoms than are 
electrons \cite{mitroy02b} (this result is based on results for 
systems with $^2$S$^e$ symmetry).  Since there is a low-energy 
$^2$P$^o$ shape resonance in $e^-$-Mg 
scattering \cite{romanyak00a,bartschat04a}, one could reasonably 
infer that the $e^+$-Mg system would also have a $^2$P$^o$ shape 
resonance or, alternatively, support a bound state.  

\begin{figure}[tbh]
\centering{
\includegraphics[width=8.7cm,angle=0]{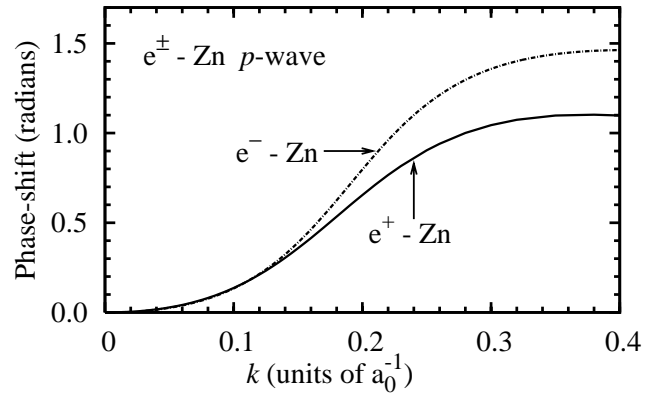}
}
\caption[]{ \label{zncompare}
The $p$-wave phase shifts as a function of $k$ (in units of 
$a_0^{-1}$) for $e^{\pm}$-Zn scattering as given by the present 
calculation and the BSRM calculation \cite{zatsarinny05a}.
}
\end{figure}

The $e^+$-Zn cross section has a broad feature in the $p$-wave 
at about 0.6 eV that could be interpreted as a resonance or a 
precursor to a resonance.  It should be noted that a similar
feature occurs in $e^-$-Zn scattering \cite{zatsarinny05a} at 
roughly the same energy.  Figure \ref{zncompare} compares the 
$p$-wave phase shift from the B-spline $R$-matrix (BSRM) 
calculation of $e^-$-Zn scattering \cite{zatsarinny05a} with 
the present $e^+$-Zn phase shift.   While the BSRM is probably 
not converged with respect to the enlargement of the channel 
space, the low-energy elastic cross section does a reasonable 
job at reproducing the electron transmission experiment of 
Burrow {\em et al} \cite{burrow76a}.  The similarity between
the electron and positron $p$-wave phase shifts for $k < 0.10$ 
$a_0^{-1}$ is expected 
since the low-energy phase shifts will be dominated by the long 
range polarization potentials.  It is interesting to speculate upon
whether the polarization potential will lead to $e^+$-Zn phase 
shifts that are larger than the $e^-$-Zn phase shifts for the 
$^2$P$^o$ symmetry, as well as the $^2$S$^e$ symmetry.  The
comparison depicted in Figure \ref{zncompare} shows that the
$e^-$-Zn $^2$P$^o$ phase shift is larger than the $e^+$-Zn 
phase shift for $k > 0.14$ $a_0^{-1}$.  This is possibly due  
to the electron seeing an attractive static potential as it
penetrates the centrifugal barrier while the positron experiences
a repulsive potential.  However, it would be best to test this
conjecture using models for the Zn target which are exactly
the same. 

The existence of the $e^+$-Mg resonance and the structure in 
$e^+$-Zn suggest that other group II atoms might support a
$^2$P$^o$ shape resonance.  The dipole polarizability of
beryllium is only a bit smaller than that of Zn, so some sort
of structure in the $^2$P$^o$ partial wave is expected.
The cadmium atom, on the other hand has a larger 
polarizability than zinc, so a more pronounced resonance
should be expected.    

The actual polarization potentials used here represent a departure
from those used in some previous calculations of positron-atom 
interactions \cite{nakanishi86a,nakanishi86b,bromley98,mitroy02a,bromley02d}.  
All of these previous works use polarization potentials with a
cut-off function that leads to a potential that is always smaller
in magnitude than that of the $\alpha_d/(2r^4)$ asymptotic potential. 
The present polarization potentials have bulges in the outer valence
region that are larger in magnitude than the asymptotic potential 
in that region.

A possible area of application of the current approach  
would be to positron-molecule scattering.  However, this would 
require improvements in the technology of positron-molecule CI 
calculations.  The best calculations so far carried out 
\cite{strasburger04a,buenker06a,buenker07a} do not treat the electron-positron 
dynamics nearly was well as the present CI calculations on atoms.   

\acknowledgments 

This work was supported under the Australian Research Council's
Discovery Program (project number 0665020).  The authors would 
like to thank Oleg Zatsarinny and Klaus Bartschat for providing 
$e^-$-Zn phase shifts in tabular form. 
The calculations upon the $e^+$Mg and $e^+$Zn systems were performed 
on GNU/Linux clusters hosted at the SDSU Computational Science Research 
Center and the South Australian Partnership for Advanced Computing 
(SAPAC).  The authors would like to thank Dr. James Otto and Grant Ward
for their system administration. 


\begin{thebibliography}{80}
\expandafter\ifx\csname natexlab\endcsname\relax\def\natexlab#1{#1}\fi
\expandafter\ifx\csname bibnamefont\endcsname\relax
  \def\bibnamefont#1{#1}\fi
\expandafter\ifx\csname bibfnamefont\endcsname\relax
  \def\bibfnamefont#1{#1}\fi
\expandafter\ifx\csname citenamefont\endcsname\relax
  \def\citenamefont#1{#1}\fi
\expandafter\ifx\csname url\endcsname\relax
  \def\url#1{\texttt{#1}}\fi
\expandafter\ifx\csname urlprefix\endcsname\relax\def\urlprefix{URL }\fi
\providecommand{\bibinfo}[2]{#2}
\providecommand{\eprint}[2][]{\url{#2}}

\bibitem[{\citenamefont{Burke and Joachain}(1995)}]{burke95}
\bibinfo{author}{\bibfnamefont{P.~G.} \bibnamefont{Burke}} \bibnamefont{and}
  \bibinfo{author}{\bibfnamefont{C.~J.} \bibnamefont{Joachain}},
  \emph{\bibinfo{title}{Theory of electron-atom collisions. Part 1 potential
  scattering}} (\bibinfo{publisher}{Plenum}, \bibinfo{address}{New York},
  \bibinfo{year}{1995}).

\bibitem[{\citenamefont{Schwartz}(1961)}]{schwartz61a}
\bibinfo{author}{\bibfnamefont{C.}~\bibnamefont{Schwartz}},
  \bibinfo{journal}{Ann.~Phys.~NY} \textbf{\bibinfo{volume}{16}},
  \bibinfo{pages}{36} (\bibinfo{year}{1961}).

\bibitem[{\citenamefont{Nesbet}(1980)}]{nesbet80a}
\bibinfo{author}{\bibfnamefont{R.~K.} \bibnamefont{Nesbet}},
  \emph{\bibinfo{title}{Variational methods in electron-atom scattering
  theory}} (\bibinfo{publisher}{Plenum}, \bibinfo{address}{New York},
  \bibinfo{year}{1980}).

\bibitem[{\citenamefont{Schneider and Rescigno}(1988)}]{schneider88a}
\bibinfo{author}{\bibfnamefont{B.~I.} \bibnamefont{Schneider}}
  \bibnamefont{and} \bibinfo{author}{\bibfnamefont{T.~N.}
  \bibnamefont{Rescigno}}, \bibinfo{journal}{Phys.~Rev.~A}
  \textbf{\bibinfo{volume}{37}}, \bibinfo{pages}{3749} (\bibinfo{year}{1988}).

\bibitem[{\citenamefont{Bromley and Mitroy}(2003)}]{bromley03a}
\bibinfo{author}{\bibfnamefont{M.~W.~J.} \bibnamefont{Bromley}}
  \bibnamefont{and} \bibinfo{author}{\bibfnamefont{J.}~\bibnamefont{Mitroy}},
  \bibinfo{journal}{Phys.~Rev.~A} \textbf{\bibinfo{volume}{67}},
  \bibinfo{pages}{062709} (\bibinfo{year}{2003}).

\bibitem[{\citenamefont{Fano and Lee}(1973)}]{fano73a}
\bibinfo{author}{\bibfnamefont{U.}~\bibnamefont{Fano}} \bibnamefont{and}
  \bibinfo{author}{\bibfnamefont{C.~M.} \bibnamefont{Lee}},
  \bibinfo{journal}{Phys.~Rev.~Lett.} \textbf{\bibinfo{volume}{31}},
  \bibinfo{pages}{1573} (\bibinfo{year}{1973}).

\bibitem[{\citenamefont{Burke et~al.}(1971)\citenamefont{Burke, Hibbert, and
  Robb}}]{burke71a}
\bibinfo{author}{\bibfnamefont{P.~G.} \bibnamefont{Burke}},
  \bibinfo{author}{\bibfnamefont{A.}~\bibnamefont{Hibbert}}, \bibnamefont{and}
  \bibinfo{author}{\bibfnamefont{W.~D.} \bibnamefont{Robb}},
  \bibinfo{journal}{J.~Phys.~B} \textbf{\bibinfo{volume}{5}},
  \bibinfo{pages}{153} (\bibinfo{year}{1971}).

\bibitem[{\citenamefont{Barrett et~al.}(1983)\citenamefont{Barrett, Robson, and
  Tobocman}}]{barrett83a}
\bibinfo{author}{\bibfnamefont{R.~F.} \bibnamefont{Barrett}},
  \bibinfo{author}{\bibfnamefont{B.~A.} \bibnamefont{Robson}},
  \bibnamefont{and} \bibinfo{author}{\bibfnamefont{W.}~\bibnamefont{Tobocman}},
  \bibinfo{journal}{Rev.~Mod.~Phys.} \textbf{\bibinfo{volume}{55}},
  \bibinfo{pages}{155} (\bibinfo{year}{1983}).

\bibitem[{\citenamefont{Bromley and Mitroy}(2006{\natexlab{a}})}]{bromley06f}
\bibinfo{author}{\bibfnamefont{M.~W.~J.} \bibnamefont{Bromley}}
  \bibnamefont{and} \bibinfo{author}{\bibfnamefont{J.}~\bibnamefont{Mitroy}},
  \bibinfo{journal}{Phys.~Rev.~Lett.} \textbf{\bibinfo{volume}{97}},
  \bibinfo{pages}{183402} (\bibinfo{year}{2006}{\natexlab{a}}).

\bibitem[{\citenamefont{McEachran et~al.}(1977)\citenamefont{McEachran, Morgan,
  Ryman, and Stauffer}}]{mceachran77}
\bibinfo{author}{\bibfnamefont{R.~P.} \bibnamefont{McEachran}},
  \bibinfo{author}{\bibfnamefont{D.~L.} \bibnamefont{Morgan}},
  \bibinfo{author}{\bibfnamefont{A.~G.} \bibnamefont{Ryman}}, \bibnamefont{and}
  \bibinfo{author}{\bibfnamefont{A.~D.} \bibnamefont{Stauffer}},
  \bibinfo{journal}{J.~Phys.~B} \textbf{\bibinfo{volume}{10}},
  \bibinfo{pages}{663} (\bibinfo{year}{1977}).

\bibitem[{\citenamefont{Higgins et~al.}(1990)\citenamefont{Higgins, Burke, and
  Walters}}]{higgins90}
\bibinfo{author}{\bibfnamefont{K.}~\bibnamefont{Higgins}},
  \bibinfo{author}{\bibfnamefont{P.~G.} \bibnamefont{Burke}}, \bibnamefont{and}
  \bibinfo{author}{\bibfnamefont{H.~R.~J.} \bibnamefont{Walters}},
  \bibinfo{journal}{J.~Phys.~B} \textbf{\bibinfo{volume}{23}},
  \bibinfo{pages}{1345} (\bibinfo{year}{1990}).

\bibitem[{\citenamefont{Strasburger and Chojnacki}(1995)}]{strasburger95}
\bibinfo{author}{\bibfnamefont{K.}~\bibnamefont{Strasburger}} \bibnamefont{and}
  \bibinfo{author}{\bibfnamefont{H.}~\bibnamefont{Chojnacki}},
  \bibinfo{journal}{Chem.~Phys.~Lett.} \textbf{\bibinfo{volume}{241}},
  \bibinfo{pages}{485} (\bibinfo{year}{1995}).

\bibitem[{\citenamefont{Schrader}(1998)}]{schrader98}
\bibinfo{author}{\bibfnamefont{D.~M.} \bibnamefont{Schrader}},
  \bibinfo{journal}{Nucl.~Instrum.~Methods~Phys.~Res.~B}
  \textbf{\bibinfo{volume}{143}}, \bibinfo{pages}{209} (\bibinfo{year}{1998}).

\bibitem[{\citenamefont{Mitroy and Ryzhikh}(1999{\natexlab{a}})}]{mitroy99c}
\bibinfo{author}{\bibfnamefont{J.}~\bibnamefont{Mitroy}} \bibnamefont{and}
  \bibinfo{author}{\bibfnamefont{G.~G.} \bibnamefont{Ryzhikh}},
  \bibinfo{journal}{J.~Phys.~B} \textbf{\bibinfo{volume}{32}},
  \bibinfo{pages}{2831} (\bibinfo{year}{1999}{\natexlab{a}}).

\bibitem[{\citenamefont{Dzuba et~al.}(1999)\citenamefont{Dzuba, Flambaum,
  Gribakin, and Harabati}}]{dzuba99}
\bibinfo{author}{\bibfnamefont{V.~A.} \bibnamefont{Dzuba}},
  \bibinfo{author}{\bibfnamefont{V.~V.} \bibnamefont{Flambaum}},
  \bibinfo{author}{\bibfnamefont{G.~F.} \bibnamefont{Gribakin}},
  \bibnamefont{and} \bibinfo{author}{\bibfnamefont{C.}~\bibnamefont{Harabati}},
  \bibinfo{journal}{Phys.~Rev.~A} \textbf{\bibinfo{volume}{60}},
  \bibinfo{pages}{3641} (\bibinfo{year}{1999}).

\bibitem[{\citenamefont{Stathopolous and {Froese
  Fischer}}(1994)}]{stathopolous94a}
\bibinfo{author}{\bibfnamefont{A.}~\bibnamefont{Stathopolous}}
  \bibnamefont{and} \bibinfo{author}{\bibfnamefont{C.}~\bibnamefont{{Froese
  Fischer}}}, \bibinfo{journal}{Comput.~Phys.~Commun.}
  \textbf{\bibinfo{volume}{79}}, \bibinfo{pages}{268} (\bibinfo{year}{1994}).

\bibitem[{\citenamefont{Saad}(2000)}]{saad00a}
\bibinfo{editor}{\bibfnamefont{Y.}~\bibnamefont{Saad}}, ed.,
  \emph{\bibinfo{title}{Iterative Methods for Sparse Linear Systems}}
  (\bibinfo{publisher}{PWS Publishing}, \bibinfo{address}{Boston},
  \bibinfo{year}{2000}).

\bibitem[{\citenamefont{Mitroy and Bromley}(2007)}]{mitroy07a}
\bibinfo{author}{\bibfnamefont{J.}~\bibnamefont{Mitroy}} \bibnamefont{and}
  \bibinfo{author}{\bibfnamefont{M.~W.~J.} \bibnamefont{Bromley}},
  \bibinfo{journal}{Phys.~Rev.~Lett.} \textbf{\bibinfo{volume}{98}},
  \bibinfo{pages}{173001} (\bibinfo{year}{2007}).

\bibitem[{\citenamefont{Mitroy and Ryzhikh}(1999{\natexlab{b}})}]{mitroy99a}
\bibinfo{author}{\bibfnamefont{J.}~\bibnamefont{Mitroy}} \bibnamefont{and}
  \bibinfo{author}{\bibfnamefont{G.~G.} \bibnamefont{Ryzhikh}},
  \bibinfo{journal}{J.~Phys.~B} \textbf{\bibinfo{volume}{32}},
  \bibinfo{pages}{1375} (\bibinfo{year}{1999}{\natexlab{b}}).

\bibitem[{\citenamefont{Bromley and Mitroy}(2002{\natexlab{a}})}]{bromley02d}
\bibinfo{author}{\bibfnamefont{M.~W.~J.} \bibnamefont{Bromley}}
  \bibnamefont{and} \bibinfo{author}{\bibfnamefont{J.}~\bibnamefont{Mitroy}},
  \bibinfo{journal}{Phys.~Rev.~A} \textbf{\bibinfo{volume}{65}},
  \bibinfo{pages}{062506} (\bibinfo{year}{2002}{\natexlab{a}}).

\bibitem[{\citenamefont{Risberg}(1956)}]{risberg56a}
\bibinfo{author}{\bibfnamefont{V.}~\bibnamefont{Risberg}},
  \bibinfo{journal}{Arch.~Math.~Naturvidenskab} \textbf{\bibinfo{volume}{53}},
  \bibinfo{pages}{1} (\bibinfo{year}{1956}).

\bibitem[{\citenamefont{Percival}(1957)}]{percival57a}
\bibinfo{author}{\bibfnamefont{I.~C.} \bibnamefont{Percival}},
  \bibinfo{journal}{Proc.~Phys.~Soc.~A} \textbf{\bibinfo{volume}{70}},
  \bibinfo{pages}{494} (\bibinfo{year}{1957}).

\bibitem[{\citenamefont{Percival}(1960)}]{percival60a}
\bibinfo{author}{\bibfnamefont{I.~C.} \bibnamefont{Percival}},
  \bibinfo{journal}{Phys.~Rev.} \textbf{\bibinfo{volume}{119}},
  \bibinfo{pages}{159} (\bibinfo{year}{1960}).

\bibitem[{\citenamefont{Alhassid and Koonin}(1984)}]{alhassid84a}
\bibinfo{author}{\bibfnamefont{Y.}~\bibnamefont{Alhassid}} \bibnamefont{and}
  \bibinfo{author}{\bibfnamefont{S.~E.} \bibnamefont{Koonin}},
  \bibinfo{journal}{Ann.~Phys.~} \textbf{\bibinfo{volume}{155}},
  \bibinfo{pages}{108} (\bibinfo{year}{1984}).

\bibitem[{\citenamefont{Carlson et~al.}(1984)\citenamefont{Carlson,
  Pandharipande, and Wiringa}}]{carlson84a}
\bibinfo{author}{\bibfnamefont{J.}~\bibnamefont{Carlson}},
  \bibinfo{author}{\bibfnamefont{V.~R.} \bibnamefont{Pandharipande}},
  \bibnamefont{and} \bibinfo{author}{\bibfnamefont{R.~B.}
  \bibnamefont{Wiringa}}, \bibinfo{journal}{Nucl.~Phys.~A}
  \textbf{\bibinfo{volume}{424}}, \bibinfo{pages}{47} (\bibinfo{year}{1984}).

\bibitem[{\citenamefont{Shumway and Ceperley}(2001)}]{shumway01a}
\bibinfo{author}{\bibfnamefont{J.}~\bibnamefont{Shumway}} \bibnamefont{and}
  \bibinfo{author}{\bibfnamefont{D.~M.} \bibnamefont{Ceperley}},
  \bibinfo{journal}{Phys. Rev. B} \textbf{\bibinfo{volume}{63}},
  \bibinfo{pages}{165209} (\bibinfo{year}{2001}).

\bibitem[{\citenamefont{Chiesa et~al.}(2002)\citenamefont{Chiesa, Mella, and
  Morosi}}]{chiesa02a}
\bibinfo{author}{\bibfnamefont{S.}~\bibnamefont{Chiesa}},
  \bibinfo{author}{\bibfnamefont{M.}~\bibnamefont{Mella}}, \bibnamefont{and}
  \bibinfo{author}{\bibfnamefont{G.}~\bibnamefont{Morosi}},
  \bibinfo{journal}{Phys.~Rev.~A} \textbf{\bibinfo{volume}{66}},
  \bibinfo{pages}{042502} (\bibinfo{year}{2002}).

\bibitem[{\citenamefont{Nollett et~al.}(2007)\citenamefont{Nollett, Pieper,
  Wiringa, Carlson, and Hale}}]{nollet07a}
\bibinfo{author}{\bibfnamefont{K.~M.} \bibnamefont{Nollett}},
  \bibinfo{author}{\bibfnamefont{S.~C.} \bibnamefont{Pieper}},
  \bibinfo{author}{\bibfnamefont{R.~B.} \bibnamefont{Wiringa}},
  \bibinfo{author}{\bibfnamefont{J.}~\bibnamefont{Carlson}}, \bibnamefont{and}
  \bibinfo{author}{\bibfnamefont{G.~M.} \bibnamefont{Hale}},
  \bibinfo{journal}{Phys.~Rev.~Lett.} \textbf{\bibinfo{volume}{99}},
  \bibinfo{pages}{022502} (\bibinfo{year}{2007}).

\bibitem[{\citenamefont{Bromley and Mitroy}(2002{\natexlab{b}})}]{bromley02a}
\bibinfo{author}{\bibfnamefont{M.~W.~J.} \bibnamefont{Bromley}}
  \bibnamefont{and} \bibinfo{author}{\bibfnamefont{J.}~\bibnamefont{Mitroy}},
  \bibinfo{journal}{Phys.~Rev.~A} \textbf{\bibinfo{volume}{65}},
  \bibinfo{pages}{012505} (\bibinfo{year}{2002}{\natexlab{b}}).

\bibitem[{\citenamefont{Stelbovics and Winata}(1990)}]{stelbovics90a}
\bibinfo{author}{\bibfnamefont{A.~T.} \bibnamefont{Stelbovics}}
  \bibnamefont{and} \bibinfo{author}{\bibfnamefont{T.}~\bibnamefont{Winata}},
  \bibinfo{journal}{Aust.~J.~Phys.} \textbf{\bibinfo{volume}{43}},
  \bibinfo{pages}{485} (\bibinfo{year}{1990}).

\bibitem[{\citenamefont{Harris}(1967)}]{harris67a}
\bibinfo{author}{\bibfnamefont{F.~E.} \bibnamefont{Harris}},
  \bibinfo{journal}{Phys.~Rev.~Lett.} \textbf{\bibinfo{volume}{19}},
  \bibinfo{pages}{173} (\bibinfo{year}{1967}).

\bibitem[{\citenamefont{Hazi and Taylor}(1970)}]{hazi70a}
\bibinfo{author}{\bibfnamefont{A.~U.} \bibnamefont{Hazi}} \bibnamefont{and}
  \bibinfo{author}{\bibfnamefont{H.~S.} \bibnamefont{Taylor}},
  \bibinfo{journal}{Phys.~Rev.~A} \textbf{\bibinfo{volume}{1}},
  \bibinfo{pages}{1109} (\bibinfo{year}{1970}).

\bibitem[{\citenamefont{Drachman and
  Houston}(1975{\natexlab{a}})}]{drachman75a}
\bibinfo{author}{\bibfnamefont{R.~J.} \bibnamefont{Drachman}} \bibnamefont{and}
  \bibinfo{author}{\bibfnamefont{S.~K.} \bibnamefont{Houston}},
  \bibinfo{journal}{Phys.~Rev.~A} \textbf{\bibinfo{volume}{12}},
  \bibinfo{pages}{885} (\bibinfo{year}{1975}{\natexlab{a}}).

\bibitem[{\citenamefont{Drachman and
  Houston}(1975{\natexlab{b}})}]{drachman76b}
\bibinfo{author}{\bibfnamefont{R.~J.} \bibnamefont{Drachman}} \bibnamefont{and}
  \bibinfo{author}{\bibfnamefont{S.~K.} \bibnamefont{Houston}},
  \bibinfo{journal}{Phys.~Rev.~A} \textbf{\bibinfo{volume}{14}},
  \bibinfo{pages}{894} (\bibinfo{year}{1975}{\natexlab{b}}).

\bibitem[{\citenamefont{Ivanov et~al.}(2001)\citenamefont{Ivanov, Mitroy, and
  Varga}}]{ivanov01b}
\bibinfo{author}{\bibfnamefont{I.~A.} \bibnamefont{Ivanov}},
  \bibinfo{author}{\bibfnamefont{J.}~\bibnamefont{Mitroy}}, \bibnamefont{and}
  \bibinfo{author}{\bibfnamefont{K.}~\bibnamefont{Varga}},
  \bibinfo{journal}{Phys.~Rev.~Lett.} \textbf{\bibinfo{volume}{87}},
  \bibinfo{pages}{063201} (\bibinfo{year}{2001}).

\bibitem[{\citenamefont{Ivanov et~al.}(2002)\citenamefont{Ivanov, Mitroy, and
  Varga}}]{ivanov02c}
\bibinfo{author}{\bibfnamefont{I.~A.} \bibnamefont{Ivanov}},
  \bibinfo{author}{\bibfnamefont{J.}~\bibnamefont{Mitroy}}, \bibnamefont{and}
  \bibinfo{author}{\bibfnamefont{K.}~\bibnamefont{Varga}},
  \bibinfo{journal}{Phys.~Rev.~A} \textbf{\bibinfo{volume}{65}},
  \bibinfo{pages}{032703} (\bibinfo{year}{2002}).

\bibitem[{\citenamefont{Bromley et~al.}(1998)\citenamefont{Bromley, Mitroy, and
  Ryzhikh}}]{bromley98}
\bibinfo{author}{\bibfnamefont{M.~W.~J.} \bibnamefont{Bromley}},
  \bibinfo{author}{\bibfnamefont{J.}~\bibnamefont{Mitroy}}, \bibnamefont{and}
  \bibinfo{author}{\bibfnamefont{G.}~\bibnamefont{Ryzhikh}},
  \bibinfo{journal}{J.~Phys.~B} \textbf{\bibinfo{volume}{31}},
  \bibinfo{pages}{4449} (\bibinfo{year}{1998}).

\bibitem[{\citenamefont{Mitroy and Ivanov}(2002)}]{mitroy02a}
\bibinfo{author}{\bibfnamefont{J.}~\bibnamefont{Mitroy}} \bibnamefont{and}
  \bibinfo{author}{\bibfnamefont{I.~A.} \bibnamefont{Ivanov}},
  \bibinfo{journal}{Phys.~Rev.~A} \textbf{\bibinfo{volume}{65}},
  \bibinfo{pages}{042705} (\bibinfo{year}{2002}).

\bibitem[{\citenamefont{Fraser}(1968)}]{fraser68a}
\bibinfo{author}{\bibfnamefont{P.~A.} \bibnamefont{Fraser}},
  \bibinfo{journal}{Adv.~At.~Mol.~Phys.} \textbf{\bibinfo{volume}{4}},
  \bibinfo{pages}{63} (\bibinfo{year}{1968}).

\bibitem[{\citenamefont{Ryzhikh and Mitroy}(2000)}]{ryzhikh00a}
\bibinfo{author}{\bibfnamefont{G.~G.} \bibnamefont{Ryzhikh}} \bibnamefont{and}
  \bibinfo{author}{\bibfnamefont{J.}~\bibnamefont{Mitroy}},
  \bibinfo{journal}{J.~Phys.~B} \textbf{\bibinfo{volume}{33}},
  \bibinfo{pages}{2229} (\bibinfo{year}{2000}).

\bibitem[{\citenamefont{Mitroy and Barbiellini}(2002)}]{mitroy02g}
\bibinfo{author}{\bibfnamefont{J.}~\bibnamefont{Mitroy}} \bibnamefont{and}
  \bibinfo{author}{\bibfnamefont{B.}~\bibnamefont{Barbiellini}},
  \bibinfo{journal}{Phys.~Rev.~B} \textbf{\bibinfo{volume}{65}},
  \bibinfo{pages}{235103} (\bibinfo{year}{2002}).

\bibitem[{\citenamefont{Mitroy}(2005)}]{mitroy05f}
\bibinfo{author}{\bibfnamefont{J.}~\bibnamefont{Mitroy}},
  \bibinfo{journal}{Phys.~Rev.~A} \textbf{\bibinfo{volume}{72}},
  \bibinfo{pages}{062707} (\bibinfo{year}{2005}).

\bibitem[{\citenamefont{Boronski and Nieminen}(1986)}]{boronski86}
\bibinfo{author}{\bibfnamefont{E.}~\bibnamefont{Boronski}} \bibnamefont{and}
  \bibinfo{author}{\bibfnamefont{R.~M.} \bibnamefont{Nieminen}},
  \bibinfo{journal}{Phys.~Rev.~B} \textbf{\bibinfo{volume}{34}},
  \bibinfo{pages}{3820} (\bibinfo{year}{1986}).

\bibitem[{\citenamefont{Puska and Nieminen}(1994)}]{puska94}
\bibinfo{author}{\bibfnamefont{M.~J.} \bibnamefont{Puska}} \bibnamefont{and}
  \bibinfo{author}{\bibfnamefont{R.~M.} \bibnamefont{Nieminen}},
  \bibinfo{journal}{Rev.~Mod.~Phys.} \textbf{\bibinfo{volume}{66}},
  \bibinfo{pages}{841} (\bibinfo{year}{1994}).

\bibitem[{\citenamefont{Barbiellini}(2001)}]{barbiellini01a}
\bibinfo{author}{\bibfnamefont{B.}~\bibnamefont{Barbiellini}}, in
  \emph{\bibinfo{booktitle}{New Directions in Antimatter Physics and
  Chemistry}}, edited by \bibinfo{editor}{\bibfnamefont{C.~M.}
  \bibnamefont{Surko}} \bibnamefont{and} \bibinfo{editor}{\bibfnamefont{F.~A.}
  \bibnamefont{Gianturco}} (\bibinfo{publisher}{Kluwer Academic Publishers},
  \bibinfo{address}{The Netherlands}, \bibinfo{year}{2001}), p.
  \bibinfo{pages}{127}.

\bibitem[{\citenamefont{Bromley and Mitroy}(2002{\natexlab{c}})}]{bromley02b}
\bibinfo{author}{\bibfnamefont{M.~W.~J.} \bibnamefont{Bromley}}
  \bibnamefont{and} \bibinfo{author}{\bibfnamefont{J.}~\bibnamefont{Mitroy}},
  \bibinfo{journal}{Phys.~Rev.~A} \textbf{\bibinfo{volume}{65}},
  \bibinfo{pages}{062505} (\bibinfo{year}{2002}{\natexlab{c}}).

\bibitem[{\citenamefont{Bromley and Mitroy}(2002{\natexlab{d}})}]{bromley02e}
\bibinfo{author}{\bibfnamefont{M.~W.~J.} \bibnamefont{Bromley}}
  \bibnamefont{and} \bibinfo{author}{\bibfnamefont{J.}~\bibnamefont{Mitroy}},
  \bibinfo{journal}{Phys.~Rev.~A} \textbf{\bibinfo{volume}{66}},
  \bibinfo{pages}{062504} (\bibinfo{year}{2002}{\natexlab{d}}).

\bibitem[{\citenamefont{Mitroy and Bromley}(2006)}]{mitroy06a}
\bibinfo{author}{\bibfnamefont{J.}~\bibnamefont{Mitroy}} \bibnamefont{and}
  \bibinfo{author}{\bibfnamefont{M.~W.~J.} \bibnamefont{Bromley}},
  \bibinfo{journal}{Phys.~Rev.~A} \textbf{\bibinfo{volume}{73}},
  \bibinfo{pages}{052712} (\bibinfo{year}{2006}).

\bibitem[{\citenamefont{Mitroy and Bromley}(2003)}]{mitroy03f}
\bibinfo{author}{\bibfnamefont{J.}~\bibnamefont{Mitroy}} \bibnamefont{and}
  \bibinfo{author}{\bibfnamefont{M.~W.~J.} \bibnamefont{Bromley}},
  \bibinfo{journal}{Phys.~Rev.~A} \textbf{\bibinfo{volume}{68}},
  \bibinfo{pages}{052714} (\bibinfo{year}{2003}).

\bibitem[{\citenamefont{Mitroy et~al.}(2002)\citenamefont{Mitroy, Bromley, and
  Ryzhikh}}]{mitroy02b}
\bibinfo{author}{\bibfnamefont{J.}~\bibnamefont{Mitroy}},
  \bibinfo{author}{\bibfnamefont{M.~W.~J.} \bibnamefont{Bromley}},
  \bibnamefont{and} \bibinfo{author}{\bibfnamefont{G.~G.}
  \bibnamefont{Ryzhikh}}, \bibinfo{journal}{J.~Phys.~B}
  \textbf{\bibinfo{volume}{35}}, \bibinfo{pages}{R81} (\bibinfo{year}{2002}).

\bibitem[{\citenamefont{Schwartz}(1962)}]{schwartz62a}
\bibinfo{author}{\bibfnamefont{C.}~\bibnamefont{Schwartz}},
  \bibinfo{journal}{Phys.~Rev.} \textbf{\bibinfo{volume}{126}},
  \bibinfo{pages}{1015} (\bibinfo{year}{1962}).

\bibitem[{\citenamefont{Carroll et~al.}(1979)\citenamefont{Carroll,
  Silverstone, and Metzger}}]{carroll79a}
\bibinfo{author}{\bibfnamefont{D.~P.} \bibnamefont{Carroll}},
  \bibinfo{author}{\bibfnamefont{H.~J.} \bibnamefont{Silverstone}},
  \bibnamefont{and} \bibinfo{author}{\bibfnamefont{R.~P.}
  \bibnamefont{Metzger}}, \bibinfo{journal}{J.~Chem.~Phys.}
  \textbf{\bibinfo{volume}{71}}, \bibinfo{pages}{4142} (\bibinfo{year}{1979}).

\bibitem[{\citenamefont{Hill}(1985)}]{hill85a}
\bibinfo{author}{\bibfnamefont{R.~N.} \bibnamefont{Hill}},
  \bibinfo{journal}{J.~Chem.~Phys.} \textbf{\bibinfo{volume}{83}},
  \bibinfo{pages}{1173} (\bibinfo{year}{1985}).

\bibitem[{\citenamefont{Salomonson and Oster}(1989)}]{salomonson89b}
\bibinfo{author}{\bibfnamefont{S.}~\bibnamefont{Salomonson}} \bibnamefont{and}
  \bibinfo{author}{\bibfnamefont{P.}~\bibnamefont{Oster}},
  \bibinfo{journal}{Phys.~Rev.~A} \textbf{\bibinfo{volume}{40}},
  \bibinfo{pages}{5559} (\bibinfo{year}{1989}).

\bibitem[{\citenamefont{Gribakin and Ludlow}(2002)}]{gribakin02a}
\bibinfo{author}{\bibfnamefont{G.~F.} \bibnamefont{Gribakin}} \bibnamefont{and}
  \bibinfo{author}{\bibfnamefont{J.}~\bibnamefont{Ludlow}},
  \bibinfo{journal}{J.~Phys.~B} \textbf{\bibinfo{volume}{35}},
  \bibinfo{pages}{339} (\bibinfo{year}{2002}).

\bibitem[{\citenamefont{Bromley and Mitroy}(2007)}]{bromley07a}
\bibinfo{author}{\bibfnamefont{M.~W.~J.} \bibnamefont{Bromley}}
  \bibnamefont{and} \bibinfo{author}{\bibfnamefont{J.}~\bibnamefont{Mitroy}},
  \bibinfo{journal}{Int.~J.~Quantum~Chem.} \textbf{\bibinfo{volume}{107}},
  \bibinfo{pages}{1150} (\bibinfo{year}{2007}).

\bibitem[{\citenamefont{Bromley and Mitroy}(2006{\natexlab{b}})}]{bromley06c}
\bibinfo{author}{\bibfnamefont{M.~W.~J.} \bibnamefont{Bromley}}
  \bibnamefont{and} \bibinfo{author}{\bibfnamefont{J.}~\bibnamefont{Mitroy}},
  \bibinfo{journal}{Phys.~Rev.~A} \textbf{\bibinfo{volume}{73}},
  \bibinfo{pages}{032507} (\bibinfo{year}{2006}{\natexlab{b}}).

\bibitem[{\citenamefont{Mitroy and Zhang}(2008)}]{mitroy08f}
\bibinfo{author}{\bibfnamefont{J.}~\bibnamefont{Mitroy}} \bibnamefont{and}
  \bibinfo{author}{\bibfnamefont{J.~Y.} \bibnamefont{Zhang}},
  \bibinfo{pages}{unpublished} (\bibinfo{year}{2008}).

\bibitem[{\citenamefont{Szmytkowski}(1993{\natexlab{a}})}]{szmytkowski93b}
\bibinfo{author}{\bibfnamefont{R.}~\bibnamefont{Szmytkowski}},
  \bibinfo{journal}{Acta~Phys.~Pol.~A} \textbf{\bibinfo{volume}{84}},
  \bibinfo{pages}{1035} (\bibinfo{year}{1993}{\natexlab{a}}).

\bibitem[{\citenamefont{Dzuba et~al.}(1995)\citenamefont{Dzuba, Flambaum,
  Gribakin, and King}}]{dzuba95b}
\bibinfo{author}{\bibfnamefont{V.~A.} \bibnamefont{Dzuba}},
  \bibinfo{author}{\bibfnamefont{V.~V.} \bibnamefont{Flambaum}},
  \bibinfo{author}{\bibfnamefont{G.~F.} \bibnamefont{Gribakin}},
  \bibnamefont{and} \bibinfo{author}{\bibfnamefont{W.~A.} \bibnamefont{King}},
  \bibinfo{journal}{Phys.~Rev.~A} \textbf{\bibinfo{volume}{52}},
  \bibinfo{pages}{4541} (\bibinfo{year}{1995}).

\bibitem[{\citenamefont{McEachran and Stauffer}(1998)}]{mceachran98}
\bibinfo{author}{\bibfnamefont{R.~P.} \bibnamefont{McEachran}}
  \bibnamefont{and} \bibinfo{author}{\bibfnamefont{A.~D.}
  \bibnamefont{Stauffer}},
  \bibinfo{journal}{Nucl.~Instrum.~Methods~Phys.~Res.~B}
  \textbf{\bibinfo{volume}{143}}, \bibinfo{pages}{199} (\bibinfo{year}{1998}).

\bibitem[{\citenamefont{Ralchenko et~al.}(2007)\citenamefont{Ralchenko, Jou,
  Kelleher, Kramida, Musgrove, Reader, Wiese, and Olsen}}]{nistasd312}
\bibinfo{author}{\bibfnamefont{Y.}~\bibnamefont{Ralchenko}},
  \bibinfo{author}{\bibfnamefont{F.-C.} \bibnamefont{Jou}},
  \bibinfo{author}{\bibfnamefont{D.}~\bibnamefont{Kelleher}},
  \bibinfo{author}{\bibfnamefont{A.}~\bibnamefont{Kramida}},
  \bibinfo{author}{\bibfnamefont{A.}~\bibnamefont{Musgrove}},
  \bibinfo{author}{\bibfnamefont{J.}~\bibnamefont{Reader}},
  \bibinfo{author}{\bibfnamefont{W.}~\bibnamefont{Wiese}}, \bibnamefont{and}
  \bibinfo{author}{\bibfnamefont{K.}~\bibnamefont{Olsen}},
  \emph{\bibinfo{title}{{NIST Atomic Spectra Database Version 3.1.2}}}
  (\bibinfo{year}{2007}),
  \urlprefix\url{http://physics.nist.gov/cgi-bin/AtData/main_asd}.

\bibitem[{\citenamefont{Goebel et~al.}(1996)\citenamefont{Goebel, Hohm, and
  Maroulis}}]{goebel96b}
\bibinfo{author}{\bibfnamefont{D.}~\bibnamefont{Goebel}},
  \bibinfo{author}{\bibfnamefont{U.}~\bibnamefont{Hohm}}, \bibnamefont{and}
  \bibinfo{author}{\bibfnamefont{G.}~\bibnamefont{Maroulis}},
  \bibinfo{journal}{Phys.~Rev.~A} \textbf{\bibinfo{volume}{54}},
  \bibinfo{pages}{1973} (\bibinfo{year}{1996}).

\bibitem[{\citenamefont{Kurtz and Jordan}(1981)}]{kurtz81}
\bibinfo{author}{\bibfnamefont{H.~A.} \bibnamefont{Kurtz}} \bibnamefont{and}
  \bibinfo{author}{\bibfnamefont{K.~D.} \bibnamefont{Jordan}},
  \bibinfo{journal}{J.~Phys.~B} \textbf{\bibinfo{volume}{14}},
  \bibinfo{pages}{4361} (\bibinfo{year}{1981}).

\bibitem[{\citenamefont{Szmytkowski}(1993{\natexlab{b}})}]{szmytkowski93a}
\bibinfo{author}{\bibfnamefont{R.}~\bibnamefont{Szmytkowski}},
  \bibinfo{journal}{J.~Phys.~II} \textbf{\bibinfo{volume}{3}},
  \bibinfo{pages}{183} (\bibinfo{year}{1993}{\natexlab{b}}).

\bibitem[{\citenamefont{Gribakin and King}(1996)}]{gribakin96}
\bibinfo{author}{\bibfnamefont{G.~F.} \bibnamefont{Gribakin}} \bibnamefont{and}
  \bibinfo{author}{\bibfnamefont{W.~A.} \bibnamefont{King}},
  \bibinfo{journal}{Can.~J.~Phys.} \textbf{\bibinfo{volume}{74}},
  \bibinfo{pages}{449} (\bibinfo{year}{1996}).

\bibitem[{\citenamefont{Campeanu et~al.}(1998)\citenamefont{Campeanu,
  McEachran, Parcell, and Stauffer}}]{campeanu98a}
\bibinfo{author}{\bibfnamefont{R.}~\bibnamefont{Campeanu}},
  \bibinfo{author}{\bibfnamefont{R.~P.} \bibnamefont{McEachran}},
  \bibinfo{author}{\bibfnamefont{L.~A.} \bibnamefont{Parcell}},
  \bibnamefont{and} \bibinfo{author}{\bibfnamefont{A.~D.}
  \bibnamefont{Stauffer}},
  \bibinfo{journal}{Nucl.~Instrum.~Methods~Phys.~Res.~B}
  \textbf{\bibinfo{volume}{143}}, \bibinfo{pages}{21} (\bibinfo{year}{1998}).

\bibitem[{\citenamefont{Peng et~al.}(2007)\citenamefont{Peng, Cheng, and
  Zhou}}]{peng07a}
\bibinfo{author}{\bibfnamefont{Y.}~\bibnamefont{Peng}},
  \bibinfo{author}{\bibfnamefont{C.}~\bibnamefont{Cheng}}, \bibnamefont{and}
  \bibinfo{author}{\bibfnamefont{Y.~J.} \bibnamefont{Zhou}},
  \bibinfo{journal}{Chin.~Phys.~Lett.} \textbf{\bibinfo{volume}{24}},
  \bibinfo{pages}{625} (\bibinfo{year}{2007}).

\bibitem[{\citenamefont{Stein et~al.}(1998)\citenamefont{Stein, Harte, Jiang,
  Kauppila, Kwan, Li, and Zhou}}]{stein98a}
\bibinfo{author}{\bibfnamefont{T.~S.} \bibnamefont{Stein}},
  \bibinfo{author}{\bibfnamefont{M.}~\bibnamefont{Harte}},
  \bibinfo{author}{\bibfnamefont{J.}~\bibnamefont{Jiang}},
  \bibinfo{author}{\bibfnamefont{W.~E.} \bibnamefont{Kauppila}},
  \bibinfo{author}{\bibfnamefont{C.~K.} \bibnamefont{Kwan}},
  \bibinfo{author}{\bibfnamefont{H.}~\bibnamefont{Li}}, \bibnamefont{and}
  \bibinfo{author}{\bibfnamefont{S.}~\bibnamefont{Zhou}},
  \bibinfo{journal}{Nucl.~Instrum.~Methods~Phys.~Res.~B}
  \textbf{\bibinfo{volume}{143}}, \bibinfo{pages}{68} (\bibinfo{year}{1998}).

\bibitem[{\citenamefont{Stein et~al.}(1996)\citenamefont{Stein, Jiang,
  Kauppila, Kwan, Li, Surdutovich, and Zhou}}]{stein96}
\bibinfo{author}{\bibfnamefont{T.~S.} \bibnamefont{Stein}},
  \bibinfo{author}{\bibfnamefont{J.}~\bibnamefont{Jiang}},
  \bibinfo{author}{\bibfnamefont{W.~E.} \bibnamefont{Kauppila}},
  \bibinfo{author}{\bibfnamefont{C.~K.} \bibnamefont{Kwan}},
  \bibinfo{author}{\bibfnamefont{H.}~\bibnamefont{Li}},
  \bibinfo{author}{\bibfnamefont{E.}~\bibnamefont{Surdutovich}},
  \bibnamefont{and} \bibinfo{author}{\bibfnamefont{S.}~\bibnamefont{Zhou}},
  \bibinfo{journal}{Can.~J.~Phys.} \textbf{\bibinfo{volume}{74}},
  \bibinfo{pages}{313} (\bibinfo{year}{1996}).

\bibitem[{\citenamefont{Surdutovich et~al.}(2003)\citenamefont{Surdutovich,
  Harte, Kauppila, Kwan, and Stein}}]{surdutovich03a}
\bibinfo{author}{\bibfnamefont{E.}~\bibnamefont{Surdutovich}},
  \bibinfo{author}{\bibfnamefont{M.}~\bibnamefont{Harte}},
  \bibinfo{author}{\bibfnamefont{W.~E.} \bibnamefont{Kauppila}},
  \bibinfo{author}{\bibfnamefont{C.~K.} \bibnamefont{Kwan}}, \bibnamefont{and}
  \bibinfo{author}{\bibfnamefont{T.~S.} \bibnamefont{Stein}},
  \bibinfo{journal}{Phys.~Rev.~A} \textbf{\bibinfo{volume}{68}},
  \bibinfo{pages}{022709} (\bibinfo{year}{2003}).

\bibitem[{\citenamefont{Romanyak et~al.}(1980)\citenamefont{Romanyak, Shpenik,
  Zhukov, and Zapesochnyi}}]{romanyak00a}
\bibinfo{author}{\bibfnamefont{N.~I.} \bibnamefont{Romanyak}},
  \bibinfo{author}{\bibfnamefont{O.~B.} \bibnamefont{Shpenik}},
  \bibinfo{author}{\bibfnamefont{A.~I.} \bibnamefont{Zhukov}},
  \bibnamefont{and} \bibinfo{author}{\bibfnamefont{I.~P.}
  \bibnamefont{Zapesochnyi}}, \bibinfo{journal}{Pis'ma.~Zh.~Tekh.~Fiz.}
  \textbf{\bibinfo{volume}{6}}, \bibinfo{pages}{877} (\bibinfo{year}{1980}).

\bibitem[{\citenamefont{Bartschat et~al.}(2004)\citenamefont{Bartschat,
  Zatsarinny, Bray, Fursa, and Stelbovics}}]{bartschat04a}
\bibinfo{author}{\bibfnamefont{K.}~\bibnamefont{Bartschat}},
  \bibinfo{author}{\bibfnamefont{O.}~\bibnamefont{Zatsarinny}},
  \bibinfo{author}{\bibfnamefont{I.}~\bibnamefont{Bray}},
  \bibinfo{author}{\bibfnamefont{D.~V.} \bibnamefont{Fursa}}, \bibnamefont{and}
  \bibinfo{author}{\bibfnamefont{A.~T.} \bibnamefont{Stelbovics}},
  \bibinfo{journal}{J.~Phys.~B} \textbf{\bibinfo{volume}{37}},
  \bibinfo{pages}{2617} (\bibinfo{year}{2004}).

\bibitem[{\citenamefont{Zatsarinny and Bartschat}(2005)}]{zatsarinny05a}
\bibinfo{author}{\bibfnamefont{O.}~\bibnamefont{Zatsarinny}} \bibnamefont{and}
  \bibinfo{author}{\bibfnamefont{K.}~\bibnamefont{Bartschat}},
  \bibinfo{journal}{Phys.~Rev.~A} \textbf{\bibinfo{volume}{71}},
  \bibinfo{pages}{022716} (\bibinfo{year}{2005}).

\bibitem[{\citenamefont{Burrow et~al.}(1976)\citenamefont{Burrow, Michejda, and
  Comer}}]{burrow76a}
\bibinfo{author}{\bibfnamefont{P.~D.} \bibnamefont{Burrow}},
  \bibinfo{author}{\bibfnamefont{J.~A.} \bibnamefont{Michejda}},
  \bibnamefont{and} \bibinfo{author}{\bibfnamefont{J.}~\bibnamefont{Comer}},
  \bibinfo{journal}{J.~Phys.~B} \textbf{\bibinfo{volume}{9}},
  \bibinfo{pages}{3225} (\bibinfo{year}{1976}).

\bibitem[{\citenamefont{Nakanishi and
  Schrader}(1986{\natexlab{a}})}]{nakanishi86a}
\bibinfo{author}{\bibfnamefont{H.}~\bibnamefont{Nakanishi}} \bibnamefont{and}
  \bibinfo{author}{\bibfnamefont{D.~M.} \bibnamefont{Schrader}},
  \bibinfo{journal}{Phys. Rev. A} \textbf{\bibinfo{volume}{34}},
  \bibinfo{pages}{1810} (\bibinfo{year}{1986}{\natexlab{a}}).

\bibitem[{\citenamefont{Nakanishi and
  Schrader}(1986{\natexlab{b}})}]{nakanishi86b}
\bibinfo{author}{\bibfnamefont{H.}~\bibnamefont{Nakanishi}} \bibnamefont{and}
  \bibinfo{author}{\bibfnamefont{D.~M.} \bibnamefont{Schrader}},
  \bibinfo{journal}{Phys. Rev. A} \textbf{\bibinfo{volume}{34}},
  \bibinfo{pages}{1823} (\bibinfo{year}{1986}{\natexlab{b}}).

\bibitem[{\citenamefont{Strasburger}(2004)}]{strasburger04a}
\bibinfo{author}{\bibfnamefont{K.}~\bibnamefont{Strasburger}},
  \bibinfo{journal}{Struct.~Chem.} \textbf{\bibinfo{volume}{15}},
  \bibinfo{pages}{415} (\bibinfo{year}{2004}).

\bibitem[{\citenamefont{Buenker et~al.}(2006)\citenamefont{Buenker, Liebermann,
  Tachikawa, Pichl, and Kimura}}]{buenker06a}
\bibinfo{author}{\bibfnamefont{R.~J.} \bibnamefont{Buenker}},
  \bibinfo{author}{\bibfnamefont{H.}~\bibnamefont{Liebermann}},
  \bibinfo{author}{\bibfnamefont{M.}~\bibnamefont{Tachikawa}},
  \bibinfo{author}{\bibfnamefont{L.}~\bibnamefont{Pichl}}, \bibnamefont{and}
  \bibinfo{author}{\bibfnamefont{M.}~\bibnamefont{Kimura}},
  \bibinfo{journal}{Nucl.~Instrum.~Methods~Phys.~Res.~B}
  \textbf{\bibinfo{volume}{247}}, \bibinfo{pages}{47} (\bibinfo{year}{2006}).

\bibitem[{\citenamefont{Buenker et~al.}(2007)\citenamefont{Buenker, Liebermann,
  Pichl, Tachikawa, and Kimura}}]{buenker07a}
\bibinfo{author}{\bibfnamefont{R.~J.} \bibnamefont{Buenker}},
  \bibinfo{author}{\bibfnamefont{H.}~\bibnamefont{Liebermann}},
  \bibinfo{author}{\bibfnamefont{L.}~\bibnamefont{Pichl}},
  \bibinfo{author}{\bibfnamefont{M.}~\bibnamefont{Tachikawa}},
  \bibnamefont{and} \bibinfo{author}{\bibfnamefont{M.}~\bibnamefont{Kimura}},
  \bibinfo{journal}{J.~Chem.~Phys.} \textbf{\bibinfo{volume}{126}},
  \bibinfo{pages}{104305} (\bibinfo{year}{2007}).

\end{thebibliography}

\end{document}